\documentclass[12pt]{article}
\usepackage[a4paper,left=2.5cm,right=2.5cm,top=3cm,bottom=3cm]{geometry}
\usepackage{amsmath,amssymb}
\usepackage{amscd}
\usepackage{graphicx}
\usepackage{epsfig}

\usepackage{amsmath, amsthm, amssymb, amsfonts}
\usepackage{setspace, graphicx, type1cm, color} 

\usepackage{slashbox,booktabs}
\makeatletter
\def\rightharpoonup@{\arrowfill@\relbar\relbar\rightharpoonup}
\newcommand{\overrightharpoonup}{\mathpalette{\overarrow@\rightharpoonup@}}
\makeatother

\renewcommand{\title}[1]{\vbox{\center\LARGE{#1}}\vspace{5mm}}
\renewcommand{\author}[1]{\vbox{\center#1}\vspace{5mm}}

\newcommand{\be}{\begin{equation}}
\newcommand{\ee}{\end{equation}}

\def\bea{\begin{eqnarray}}
\def\eea{\end{eqnarray}}

\numberwithin{equation}{section}

\newcommand{\nn}{\nonumber }

\newcommand{\commt}[2]{\left[#1, #2 \right] }

\newcommand{\bZ}{\mathbb{Z} }
\newcommand{\bC}{\mathbb{C} }

\newcommand{\ket}[1]{\left| \left. #1 \right\rangle \right.}

\newcommand{\ph}[1]{\phantom{#1}}
\newcommand{\vev}[1]{\left< #1 \right> }

\allowdisplaybreaks

\begin{document}
   \fontsize{13pt}{18pt}\selectfont 
%%%%  For title page %%%%%

\title{\bf To See Symmetry in a Forest of Trees}

\author{
 Chuan-Tsung Chan$^{ab\spadesuit}$,
         Shoichi Kawamoto$^{ab\clubsuit}$, 
         and
 	Dan Tomino$^{a\heartsuit}$\bigskip
 	\\
 	\footnotesize\it $^a$ Department of Physics, Tunghai University,
 Taichung 40704, Taiwan\\
 	\footnotesize\it $^b$ National Center for Theoretical Sciences, Hsinchu 30013, Taiwan\\
 	\footnotesize\tt
$^\spadesuit$ ctchan@go.thu.edu.tw
 	\smallskip\\
 $^\clubsuit$ kawamoto@thu.edu.tw, kawamoto@yukawa.kyoto-u.ac.jp
 	\smallskip\\
 	$^\heartsuit$ dantomino@thu.edu.tw
\footnotesize\it
	}

\date{\today}

%\maketitle

\bigskip

\begin{abstract}

\fontsize{13pt}{18pt}\selectfont 
The exact symmetry identities among four-point tree-level
amplitudes 
of bosonic open string theory
as derived by G. W. Moore are re-examined. The main
focuses of this work are: 
(1) Explicit construction of kinematic configurations and a new
polarization basis for the scattering processes. These setups
simplify greatly
the functional forms of 
the exact symmetry identities, and help us to extract easily high-energy
limits of stringy amplitudes appearing in the exact identities.
(2) Connection and comparison between D. J. Gross's
high-energy stringy symmetry and 
the exact symmetry identities as derived by G. W. Moore.
(3) Observation of symmetry patterns of stringy amplitudes with respect to the order of energy dependence in scattering amplitudes.

\end{abstract}

\vfill
\newpage
\tableofcontents
\newpage

%%% From here, spacing between the paragraphs are set
\setlength{\parskip}{1ex plus 0.5ex minus 0.2ex}

%%%%%%%% BODY %%%%%%%%%%%%%
\setcounter{footnote}{0}
%%%%%%%%%%%%%%%%%%%%%%%%%
%%%%%%%%%%%%%%%%%%%%%%%%%%%%%%%%%
%1
\section{Introduction}

   Given many tremendous progresses and miraculous achievements, string theory as we know it today is still a beautiful work under construction \cite{Green:1987sp,Polchinski:1998rq}. 
While the lack of a full non-perturbative background-independent  definition \cite{Witten:1992qy, Sen:1990hh} may await for an unexpected breakthrough, the current formulation 
does not follow the wisdom of previous paradigms such as Einstein's general theory of relativity or the standard model of the particle physics. 
Nevertheless, one can hardly imagine that a symmetry principle would be irrelevant under a proper formulation of string theory. 
To this end, many people have addressed this issue over past decades. See, for example, \cite{Witten:1988sy}.
 Notable examples are the high-energy symmetry as proposed by
D. J. Gross \cite{Gross:1988ue,Mende:1989wt} and the exact symmetry identities derived by 
G. W. Moore \cite{Moore:1993ns}.
 In the former context, one observes linear relations among high-energy scattering amplitudes among stringy excitations at the same level, and inter-level 
symmetry patterns for leading four-point tree-level amplitudes 
were
proposed in \cite{Chan:2003ee}.
 In the latter context, based on a well-defined algebraic algorithm
(described by a bracket operation defined in Section \ref{sec:2}),
one establishes
exact identities among inter-level stringy scattering amplitudes.

The basic idea behind high-energy symmetry as envisioned by Gross et
al. \cite{Gross:1988ue} is to view string theory as a higher-spin 
gauge theory with spontaneous symmetry breakdown.
Here all higher-level stringy excitations gain their masses through a
higher-spin generalization of Higgs mechanism \cite{Witten:1988sy}.
 Furthermore, if we combine both the master 
formula \cite{Chan:2003ee} of tree-level stringy amplitudes for all
transverse-polarized highest spin states at given mass levels,
together with the linear relations among leading high-energy
scattering amplitudes, these patterns strongly suggest an underlying
structure of string theory as that of equivalent theorem in the
electroweak theory \cite{Cornwall:1974km}.
The fact that we are able to deduce linear relations among the stringy
scattering amplitudes \cite{Chan:2003ee} provides further evidences
that the high-energy symmetry is a reflection of global symmetry
associated with the would-be Goldstone particles in the unbroken phase
of string theory.

 In contrast, the advantages of the identities of Moore are: 
(1) the derivation of the identities is based on a clear algebraic structure of symmetry, some subsectors of the bracket states and their bracket relations,
 based on the bracket operation, 
can be described in explicit mathematical frameworks, e.g. \cite{Lian:1992mn}.   
 (2) While there are infinitely many exact relations one can write down based on bracket algebra, these are not totally independent identities.
 In fact, one of the special features of these exact symmetry relations is that, it almost realizes a ``bootstrap'' scenario which allows us to derive infinite many 
scattering amplitudes among massive stringy excitations based on the Veneziano amplitude. 
(3) These are exact identities among stringy amplitudes, no special kinematic limits are taken (either high-energy \cite{Chan:2004yz} or Regge limits \cite{Ko:2008ft}).

    On the contrary, to display the contents of these exact identities, especially in terms of all explicit kinematics (momenta, polarizations) are definitely not a trivial task. 
There are several immediate issues which demand special efforts. 
For instance: (1) What are the physical characteristics (momenta, polarizations) of a state generated by bracket computation (referred to as a bracket state henceforth)?
 (2) From the structure of bracket computation (to be reviewed in Section \ref{sec:2}),
 it is clear that the exact identities generally relate stringy excited states at different levels. In order to view all stringy states and 
their scattering amplitudes as representations of a huge symmetry underlying string theory, it is natural to ask if bracket states generate full stringy spectrum? 
(3) Note that due to the ``dressing'' of the deformer to various seed operators  (to be defined in Section \ref{sec:2}), the stringy amplitudes as related by the exact symmetry identities 
in general have different kinematic configurations. 
Specifically, all amplitudes involved in a given symmetry identity describe  different scattering processes, in which participant particles may have different spins and momenta. 
In application to four-point scattering amplitudes, for example,
while we know that the explicit form of any Lorentz invariant four-point amplitudes must be a function of two Mandelstam variables $(s,t)$,
 there is no guarantee that all four four-point amplitudes appearing in an exact symmetry identity share the same set of Mandelstam variables. 
  
    In view of these, the symmetry identities as derived from the bracket algebra not only are generic inter-level symmetry relations but also connect amplitudes with 
different kinematic configurations.
 Clearly, these relations are based on a well-defined infinite-dimensional symmetry algebra and may cover a wider energy region as compared with, say, high-energy 
symmetries \` a la Gross.
 Nevertheless, from a physicist's point of view, if we believe that all scattering amplitudes form representations of the grand symmetry of string theory, one would 
like to have explicit actions on the scattering amplitudes as explicit functions of Mandelstam variables. 
  Indeed, while neither symmetry relations mentioned above cover complete patterns of the stringy amplitudes, 
it is still of interest and importance to see if we can make connections between these two approaches. To achieve this goal, we need to pin down the explicit kinematic dependence of the exact 
identities and study their high-energy behaviors.
 In this paper, we begin the first exploration of such a
 connection/unification based on a couple of case studies.
We make a detailed comparison of the spectrum of stringy scattering
states as generated from bracket algebra and identify a new kinematic
basis for the decomposition of the polarization tensors. 
Most importantly, through the choice of proper basis of the string state Fock 
space (Verma module), we obtain much-simplified representations of the
exact identities which allow us to extract high-energy limits easily.
Though we have worked out two specific cases,
they already provide a couple of
essential and generic features of these exact/high-energy relations.
We therefore believe that the study worked out here will be a good
starting point toward more general understanding of Moore's relation.

   This paper is organized as follows: 
we first give a brief review of Moore's derivation of exact identities
among string amplitudes in Section \ref{sec:2}.
Then we discuss the condition of conformal invariance 
on the bracket states and study their spectrum in Section \ref{sec:3}.
 In order to examine the explicit kinematic dependence of the stringy amplitudes, we give a detailed study of the 4-point kinematic configuration 
in Section \ref{sec:4}.
Here we also construct a new basis set for the helicity vector/tensor ($q$-orthonormal basis) which leads to improved expressions of exact identities. 
In Section \ref{sec:change_basis}, 
we investigate how the physical bracket states
are related to the conventional positive-norm states
as well as light-cone like physical states based on Del Giudice, Di
Vecchia, and Fubini (DDF) operators \cite{Del Giudice:1971fp,
  Green:1987sp} (referred to as DDF states).
It is also discussed that the derivations of the exact symmetry identities of the stringy tree amplitudes with explicit  kinematic dependence. 
 Two explicit cases are used to illustrate our idea. 
Finally, based on the explicit constructions, we study the high-energy expansions
of the exact identities and compare them with previous work in
Section \ref{sec:HE_expansion}. 
Section \ref{sec:conclusion} consists of summary and the discussion of future directions related to this work. 

To streamline our discussion, we only use simple examples in the main
text for various explanations. Some technical details and
further illustrative examples are collected in the appendixes for reference.
Appendix \ref{sec:bracket-operators}
aims at supplying  discussion about necessary and sufficient
conditions to make bracket operators conformally invariant.
We give a simple explanation and useful formulas of DDF states in
Appendix \ref{sec:ddf-states}.
Finally, we discuss some subtleties regarding the choice of
reference kinematic variables in the study of high-energy limits of
Moore's exact identities in Appendix \ref{sec:fixed-angle}.

\section{A brief review of G.~W.~Moore's derivation}
\label{sec:2}

\subsection{Outline of the basic idea}

Let us first begin with a brief review of the argument of
Moore \cite{Moore:1993ns}.
Throughout this paper we use convention $X^{\mu}(w)X^{\nu}(z)\sim -2\alpha'\eta^{\mu\nu}\log(w-z)$ for string world-sheet propagators.
Starting with dimension $1$ chiral currents
$J(q,w)$ (referred to as the deformer) and $V(k,z)$ (referred to as the seed operator) which carry the momenta
$q^\mu$ and $k^\mu$ respectively,
we define a new operator by
\begin{align}
\label{eq:11}
V^\text{br}(\widetilde{k},z) =
\{ \mathcal{J}(q) , V(k,z) \} \equiv
\oint_z \frac{dw}{2\pi i } \, J(q,w) V(k, z) \,,
\end{align}
where ${\cal J}(q)$ is the integrated operator of the current $J(q,w)$,
and $\widetilde{k}=k+q$ is a deformed momentum associated with this
new operator which we call a bracket operator.
This expression is well-defined when $\mathcal{J}(q) $ and $V(k,z)$
are mutually local, $2\alpha' q \cdot k \in \bZ$.
When $J(q,w) $ and $V(k,z)$ are primary, it is easy to see that
$V^\text{br}(\widetilde{k},z)$ is also primary
and defines a physical vertex operator.
We shall revisit physical state conditions later, and
temporarily we assume that both $V(k,z)$ and $V^\text{br}(\widetilde{k},z)$ are
primary.

Now we look at four-point tree-level amplitudes of bosonic open string
theory,
\begin{align}
  \label{eq:2}
  \mathcal{A}[\{ {\cal V}_i(k_i) \}] =& \int_{0}^1 dx \vev{
V_1(k_1,x) \, V_2(k_2,0) \, V_3(k_3,1) \, V_4(k_4, \infty)}
\,,
\end{align}
where ${\cal V}_i(k_i)$ are again integrated vertex operators.
On the right hand side, we did not write explicitly the ghost part which should be
understood in a standard way.
It should be noted that the string scattering amplitude
includes integration over the
other domains, $-\infty<x<0$ and $1<x<\infty$, and also the contribution
from the different ordering
of the vertex operators.
However, the relations among the scattering amplitudes we deal with in this paper are already manifest in this part, as we will see,
 so we concentrate on
this part of the scattering amplitudes.
The scattering amplitude is given as a function of independent momentum
invariants, $k_i \cdot k_j$, which we choose as the standard Mandelstam variables,
\begin{align}
  \label{eq:33}
  s=-(k_1+k_2)^2 \,,
\qquad
t=-(k_1+k_3)^2 \,,
\end{align}
for the four-point scattering amplitudes, as well as an independent set of
polarization invariants, such as $\zeta_i \cdot k_j$ or $\zeta_i \cdot \zeta_j$.

Now let us turn to unintegrated correlation functions with deformed
operators.
We may consider the second operator at $z=0$ to be deformed by the action of $\mathcal{J}(q) $ operator,
\begin{align}
  \label{eq:3}
  \vev{ V_1(k_1,x) \, \{ \mathcal{J}(q), V_2(k_2,0) \}   \, V_3(k_3,1) \  V_4(k_4, \infty) }
\,.
\end{align}
It should be noted that the momentum
conservation condition now includes $q^\mu$,
\begin{align}
  \label{eq:34}
q^\mu +  \sum_{i=1}^4 k_i^\mu =0 \,.
\end{align}
By deforming the contour
and integrating $x$ from $0$ to $1$,
we obtain a relation among scattering amplitudes,
\begin{align}
  \label{eq:35}
0=&\mathcal{A}[{\cal V}_1(k_1) {\cal V}^\text{br}_2(\widetilde{k}_2){\cal V}_3(k_3){\cal V}_4(k_4)]
\nn \\& 
+(-1)^{2\alpha'q\cdot k_1}
\mathcal{A}[{\cal V}^\text{br}_1(\widetilde{k}_1) {\cal V}_2(k_2){\cal V}_3(k_3){\cal V}_4(k_4)]
\nn \\&
+(-1)^{2\alpha' q\cdot (k_1+k_3)}\mathcal{A}[{\cal V}_1(k_1){\cal V}_2(k_2) {\cal V}^\text{br}_3(\widetilde{k}_3){\cal V}_4(k_4)]
\nn \\&
+(-1)^{2\alpha' q\cdot (k_1+k_3+k_4)}\mathcal{A}[{\cal V}_1(k_1){\cal V}_2(k_2) {\cal V}_3(k_3){\cal V}^\text{br}_4(\widetilde{k}_4)]
\,.
\end{align}
Recall that the position of the unintegrated
vertex operator $V_1(k_1,x)$ is $0< x <1<\infty$.
For this procedure to be well-defined, all $2\alpha' q \cdot k_i$ have to be
integral, while each pair of $k_i$ and $k_j$ does not need to be mutually local.
In \cite{Moore:1993ns}, this relation has been employed to derive
functional equations
for scattering amplitudes, which turn out to suffice for determining
tachyon scattering amplitudes up to a constant.
This is an intriguing result, but in this article we revisit
these relations from the viewpoint of 
the standard scattering amplitudes in the center-of-momentum frame
and their linear relations at  high energy.

To make this point clearer, we look closer to the deformation of vertex 
operators and also
recall a (fixed-angle) high-energy limit
in the string scattering amplitudes.
For four-point scattering amplitudes, we may prepare momenta $k_i$
and an extra momentum $q^\mu$ to
satisfy the following
\begin{align}
  \label{eq:36}
  k_i^2 = - m_i^2 \,,
\qquad  q^2 = -m_q^2 \,,
\qquad
2\alpha' q \cdot k_i = n_i \,,
\quad (n_i \in \bZ ) \,,
\end{align}
where the momentum conservation condition \eqref{eq:34} is also imposed.
These conditions lead to a consistency condition,
\begin{align}
  \label{eq:39}
  \sum_{i=1}^4 n_i =& 2\alpha' m_q^2 \,.
\end{align}
The deformed momenta satisfy mass-shell conditions,
\begin{align}
  \label{eq:38}
 \alpha' \widetilde{k}_i^2 =&
\alpha' (k_i+q)^2 = -\alpha' m_i^2 +n_i - \alpha' m_q^2
\equiv -\alpha'  \widetilde{m}_i^2 \,,
\end{align}
where $\alpha' \widetilde{m}_i^2$ are again integers.
Therefore the level of the vertex operator $V_i(k)$
is shifted by $\alpha' m_q^2 - n_i$.
It is easy to see from the bracket computation that if the deformed mass $\alpha' \widetilde{m}_i^2 \leq -2$,
the deformed operator identically vanishes.
The Mandelstam variables for the physical momenta 
in the second amplitude in \eqref{eq:35}
are defined as
\begin{align}
     s & \equiv - (\widetilde{k}_1 + k_2)^2
        = -(k_1 + k_2)^2 - \dfrac{1}{2 \alpha'} (n_1 + n_2) + m_q^2 \,,
       \nn\\
     t & \equiv
  -(\widetilde{k}_1 + k_3)^2
   = -(k_1 + k_3)^2 - \dfrac{1}{2 \alpha'} (n_1
       + n_3) + m_q^2 \,,
\label{sandt}
\end{align}
Following similar definitions, we obtain the relations between Mandelstam variables in various scattering amplitudes related by a exact symmetry identity.

Now we look at, for example, the second amplitude in \eqref{eq:35},
\begin{align}
  \label{eq:40}
  \mathcal{A}[{\cal V}_1^\text{br}(\widetilde{k}_1) {\cal V}_2(k_2)
  {\cal V}_3(k_3){\cal V}_4(k_4)] \,.
\end{align}
If we take a high-energy limit\footnote{%
In the high-energy regime, the string scattering amplitudes are extremely
soft and damped exponentially.
In this paper, we compare the high-energy limit of the amplitudes
up to a common exponentially damping part.
See \eqref{eq:2000_sc_general} in
Section \ref{sec:bracket-states-terms-2000-1100}.}, $\alpha' s \rightarrow \infty$ with
$t/s$ fixed, then each component of each momentum also goes to infinity;
for example,
$\widetilde{k}_1^0, |\vec{\widetilde{k}}_1| \rightarrow \infty$,
where $\vec{\widetilde{k}}_1$ is the spatial part of $26$-momentum $\widetilde{k}_1^\mu$.
The same is true for the other momenta.
On the other hand, the inner products of  $q^\mu$ with these momenta,
and also itself, are all constant.
Therefore, each component of $q^\mu$ is $\mathcal{O}(1)$ or less.
This means that the deformation due to $q^\mu$ becomes negligible
in the high-energy limit, and we can obtain a high-energy relation among
the usual scattering amplitudes with the same external momenta.
The purpose of this article is to make this observation more precise,
and demonstrate how the high-energy relations are obtained by use of a
couple of concrete examples.
We shall examine the bracket relation in terms of conventional
scattering amplitudes,
and also explore the relation to the amplitudes
based on Del Giudice, Di Vecchia and Fubini (DDF) operators.
The DDF operators are spectrum generating operators in string theory,
and play an important role, for example, in proving the no-ghost theorem.
The DDF states spanned by the action of the DDF operators thus form
a convenient basis of the positive norm states.
The DDF amplitudes, associated with these DDF states,
have an advantage that the patterns of the energy hierarchy are
much more transparent than those of conventional amplitudes.
They therefore prove to be a particularly convenient basis
when we discuss high-energy asymptotic relations among 
scattering amplitudes \cite{Chan:2005ji, Ho:2006zu, DDF-HE},
as we will briefly explain in Sec.~\ref{sec:high-energy-linear}.

\subsection{Mass and level parameters in our case studies}

In the following sections, we investigate the properties
of the bracket states and kinematics.
We shall first write down the expressions of vertex operators
\begin{align}
  \label{eq:10}
  V_{(3)}({k},z;\zeta)=&
:\left[
\frac{-i \zeta_{\mu\nu\rho}}{(2\alpha')^{3/2}} 
 \partial X^\mu\partial X^\nu \partial X^\rho 
-\frac{\zeta_{\mu ; \nu}}{2\alpha'} \partial^2 X^\mu \partial X^\nu
+\frac{i \zeta_\mu \partial^3 X^\mu}{2\sqrt{2\alpha'}}
\right] e^{i{k} \cdot X} :(z)
\,,
\\
V_{(2)}({k},z; \zeta)=&
: \bigg(
\frac{-\zeta_{\mu\nu}}{2\alpha'} \partial X^\mu \partial X^\nu
+\frac{i}{\sqrt{2\alpha'}} \zeta_\mu \partial^2 X^\mu \bigg)
 e^{i {k} \cdot X}:(z) \,,\\
  V_{(1)} (k,z; \zeta) =& \frac{i \zeta \cdot \partial X}{\sqrt{2\alpha'}}
 e^{ik \cdot X}(z)
\,, \qquad
V_{(0)}(k,z)= : e^{i k \cdot X}: (z) \,,
\end{align}
where the subscript of ${V}_{(\ell)}(k_i)$,
 $(\ell)$, denotes the level of the vertex operator;
$(0)$ is for a tachyon, $(1)$ for a massless state, and so on.
In the argument of $V_{(n)}(k,z;\zeta)$, $\zeta$ 
schematically stands for the set of
polarization tensors.
A deformer at level $n$ is represented by 
$J_{(n)}(q,w) = V_{(n)}(q,w;\zeta_q)$, and
its polarization tensors are usually denoted as $\zeta_q$
otherwise specified.
As in \eqref{eq:11}, bracket operators are written with the
superscript ``br,'' $V^\text{br}_{(n)}(\tilde{k},z; \tilde\zeta)$.
The deformation of bracket operation appears as a special form
of the polarization tensors (as well as the shift of the
momentum by $q$),  as we are about to see.

We will mainly consider the following example,
\begin{align}
%  \label{2and11-2}
m_1^2=m_q^2=0 \,,
\quad
m_2^2=m_3^2=m_4^2=-1/\alpha' \,,
\quad
n_1=n_2=-1 \,,
\quad
n_3=n_4=1 \,,\nn
\end{align}
which implies $\widetilde{m}_1^2=1/\alpha'$ and
$\widetilde{m}_2^2=0$.
It also gives
 $\widetilde{n}_1=2\alpha' \widetilde{k}_1 \cdot q = -1$
 and $\widetilde{n}_2=2\alpha' \widetilde{k}_2 \cdot q =-1$.
Namely, we prepare the following deformer operator $J_{(1)} (q,w)$
and seed operators, 
$V_{(1)} (k_1,z;\zeta_1)$ and $V_{(0)}(k_i,z)$ ($i=2,3,4$).
This choice of the parameters leads to
\begin{align}
\label{2and11}
  \mathcal{A}[{\cal V}_{(2)}^{\text{br}}(\widetilde{k}_1) {\cal
    V}_{(0)}(k_2) {\cal V}_{(0)}(k_3) {\cal V}_{(0)}(k_4)]
= \mathcal{A}[{\cal V}_{(1)}({k}_1) {\cal
  V}_{(1)}^{\text{br}}(\widetilde{k}_2) {\cal V}_{(0)}(k_3) {\cal
  V}_{(0)}(k_4)]
\,.
\end{align}
Since $\widetilde{m}_3^2=\widetilde{m}_4^2=-2/\alpha'$, the
corresponding operators identically vanish, and the relation involves
only these two amplitudes.
The explicit expressions of the 
polarization tensors of the bracket operators,
$V_{(2)}^\text{br}(\tilde{k}_1,z; \zeta^{(2)})$
and $V_{(1)}^\text{br} (\tilde{k}_2,z; \zeta_R)$
in terms of the seed and the deformer
are (for simplicity, $\alpha'=1/2$ in these expressions)
\begin{align}
  \zeta^{(2)}_{\mu\nu}(\zeta_1, \zeta_q) =&
(\zeta_q \cdot k_1)q_{(\mu} \zeta_{1\nu)}
-(\zeta_1 \cdot q)q_{(\mu} \zeta_{q\nu)}
+\zeta_{q(\mu} \zeta_{1\nu)}
\nn\\&
+\frac{1}{2}\big((\zeta_q\cdot \zeta_1)-(\zeta_q\cdot k_1)(\zeta_1 \cdot q) \big)
 q_{\mu} q_{\nu}
\label{eq:26}
\,,\\
\zeta^{(2)}_\mu(\zeta_1, \zeta_q) = &
- (\zeta_1 \cdot q) \zeta_{q\mu}
+\frac{1}{2}\big((\zeta_q\cdot \zeta_1)-(\zeta_q\cdot k_1)(\zeta_1 \cdot q) \big) q_{\mu}
\,,
\label{eq:27}
 \\
\zeta_{R\mu}(\zeta_q) =& (\zeta_q \cdot k_2) q_\mu + \zeta_{q\mu} \,.
\label{eq:28}
\end{align}
In the relation,
we call the left hand side
$\mathcal{A}[\tilde{2}000]$ amplitude and the right hand side
$\mathcal{A}[1\tilde{1}00]$
by using the sequences of the levels.
The tilde for the level number stands for deformed (bracket) operators.
In later sections, we frequently refer to this example as ``Case study I: ${\cal A}[\tilde{2}000]={\cal A}[1\tilde{1}00]$.''
The Mandelstam variables for $\mathcal{A}[\tilde{2}000]$ side
are defined by $s_{[\tilde{2}000]} = -(\tilde{k}_1+k_2)^2$
and $t_{[\tilde{2}000]} = -(\tilde{k}_1+k_3)^2$.
On the other hand, on $\mathcal{A}[1\tilde{1}00]$ side,
they are given as $s_{[1\tilde{1}00]} = -(k_1+\tilde{k}_2)^2$
and $t_{[1\tilde{1}00]} = -(k_1+k_3)^2$.
Using the mass-shell conditions and the values of $q \cdot k_i$,
one can find that these two variables are equivalent,
$s_{[\tilde{2}000]}=s_{[1\tilde{1}00]}$ and
$t_{[\tilde{2}000]}=t_{[1\tilde{1}00]}$.
We therefore simply write them as $s$ and $t$,
and the relation between the amplitudes is
understood as the relation of functions
of these $s$ and $t$,
$\mathcal{A}[\tilde{2}000](s,t) = \mathcal{A}[1\tilde{1}00](s,t)$.

In another example we will consider, we prepare a massive deformer
operator ${J}_{(2)}(q,w)$,
and the same set of seed operators,
$V_{(1)} (k_1,z;\zeta_1)$ and $V_{(0)}(k_i,z)$.
With the following choice of the parameters,
\begin{align}
%\label{3and12-2}
m_1^2=0 \,, \quad
m_q^2=1/\alpha' \,,
\quad
m_2^2=m_3^2=m_4^2=-1/\alpha' \,,
\quad
n_1=n_2=-1 \,,
\quad
n_3=n_4=2 \,,\nn
\end{align}
we obtain the following relation,
\begin{align}
\label{3and12}
  \mathcal{A}[{\cal V}_{(3)}^{\text{br}}(\widetilde{k}_1) {\cal V}_{(0)}(k_2)
  {\cal V}_{(0)}(k_3) {\cal V}_{(0)}(k_4)]
= \mathcal{A}[{\cal V}_1({k}_1) {\cal
  V}_{(2)}^{\text{br}}(\widetilde{k}_2) {\cal V}_{(0)}(k_3) {\cal V}_{(0)}(k_4)]
\,.
\end{align}
This example will be referred to as ``Case study I{}I: ${\cal
  A}[\tilde{3}000]={\cal A}[1\tilde{2}00]$''.
Note that $\widetilde{m}_1^2=2/\alpha'$, $\widetilde{n}_1=-3$,
$\widetilde{m}_2^2=1/\alpha'$ and $\widetilde{n}_2=-3$.
The explicit forms of the polarization tensors
of $V_{(3)}^\text{br}(\tilde{k}_1,z;\zeta^{(3)})$
and $V_{(2)}^\text{br}(\tilde{k}_2,z;\zeta_R)$ are spelled out as
(again $\alpha'=1/2$)
\begin{align}
\zeta^{(3)}_{\mu\nu\rho}(\zeta_q,\tilde\zeta_q; \zeta_1) 
=&
\zeta_{q(\mu\nu} \zeta_{1\rho)}
+2  q_{(\mu} \zeta_{1\nu} (\zeta_q \cdot k_1)_{\rho)}
-(\zeta_1 \cdot q) \big[
   \zeta_{q(\mu\nu} q_{\rho)}
   + (\zeta_q \cdot k_1)_{(\mu} q_\nu q_{\rho)}
   \big]
\nn\\&
+q_{(\mu} q_\nu (\zeta_q \cdot \zeta_1)_{\rho)}
+\frac{1}{2} 
\Xi_1 \,  q_{(\mu} q_\nu \zeta_{1\rho)}
-\frac{1}{6} 
\Xi_2 \, q_\mu q_\nu q_\rho
\label{eq:3000-zeta_1}
\,,\\
\zeta^{(3)}_{\mu ; \nu}(\zeta_q,\tilde\zeta_q; \zeta_1) 
=&
\tilde\zeta_{q\mu} \zeta_{1\nu}
+ 2 (\zeta_q \cdot k_1)_\mu \zeta_{1\nu}
\nn\\&
-(\zeta_1 \cdot q) \big[  2 \zeta_{q\mu\nu}
+2 (\zeta_q \cdot k_1)_\mu q_\nu
+  q_\mu (\zeta_q \cdot k_1)_\nu 
+\tilde\zeta_{q\mu} q_\nu \big]
\nn\\&
+2 (\zeta_q \cdot \zeta_1)_\mu q_\nu
+ q_\mu (\zeta_q \cdot \zeta_1)_\nu
+\frac{1}{2} \Xi_1 \,  q_\mu \zeta_{1\nu}
-\frac{1}{2} 
\Xi_2 \,  q_\mu q_\nu
\label{eq:3000-zeta_2}
\,,\\
\zeta^{(3)}_{\mu}(\zeta_q,\tilde\zeta_q; \zeta_1) 
=&
2(\zeta_q \cdot \zeta_1)_\mu
-2 (\zeta_1 \cdot q) (\zeta_q \cdot k_1)_\mu
-2(\zeta_1 \cdot q) \tilde\zeta_{q\mu}
-\frac{1}{3} \Xi_2 \,  q_\mu
\,,
\label{eq:3000-zeta_3}
\\
\Xi_1 =&   (k_1 \cdot \zeta_q \cdot k_1)
- (\tilde\zeta_q \cdot k_1) \,,\nn\\
\Xi_2 =&
 (k_1 \cdot \zeta_q \cdot k_1)(\zeta_1 \cdot q)
- (\tilde\zeta_q \cdot k_1)(\zeta_1 \cdot q)
%\nn\\&\hskip3em
-2 (\zeta_1 \cdot \zeta_q \cdot k_1)
+2  (\tilde\zeta_q \cdot \zeta_1) \,,\nn
\end{align}
\begin{align}
\zeta_{R\mu\nu}(\zeta_q, \tilde\zeta_q)
=&\zeta_{q\mu\nu} + 2\,q_{(\mu} \zeta_{q\nu)\rho} k_2^\rho
+\frac{1}{2}
\left(
k_2 \cdot \zeta_q \cdot k_2 -\tilde\zeta_q \cdot k_2 
\right)
q_\mu q_\nu
\label{eq:1200_zetaR_1}
\,,\\
\zeta_{R\mu}(\zeta_q, \tilde\zeta_q)=&
\tilde\zeta_{q\mu} + 2\zeta_{q\mu\nu}k_2^\nu
+\frac{1}{2}
\left(
k_2 \cdot \zeta_q \cdot k_2 - \tilde\zeta_q \cdot k_2 
\right)
q_\mu
\label{eq:1200_zetaR_2}
\,.
\end{align}
Here, the polarization tensors of the deformer are represented
as $\zeta_{q\mu\nu}$ and $\tilde\zeta_{q\mu}$ for distinction.
One can check that the Mandelstam variables of both hands sides
coincide also in this case, and
we write them as $s$ and $t$.

%3
\section{The bracket operators and the spectrum analysis}
\label{sec:3}

As we have explained in Section \ref{sec:2}, the basic idea underlying the derivation of exact identities among $n$-point scattering amplitudes is to 
deform the contour of the bracket operator in a null $n+1$ point scattering amplitude into separate ``dressing'' of the $n$ individual seed vertex 
operators. The bracket algebra leads to a relation among $n$ $n$-point scattering amplitudes, where each amplitude includes one deformed 
operator and other $n-1$ seed operators. While it is natural to demand that all seed operators are conformal invariant, the nature of the bracket 
operator requires some explanations. In particular, we will examine
the following questions in this section:
\begin{itemize}
\item What are necessary and sufficient conditions for the bracket operators to be conformal invariant?
\item Do bracket operators at a fixed level generate the complete positive-norm spectrum?
\end{itemize}

%3.1
\subsection{Conformal invariance of the bracket operators}
\label{confinv}
In order for the relation \eqref{eq:35} to make sense as a relation among
string scattering amplitudes, the deformed operator $V^\text{br}$ has to be
a decent vertex operator.
If $J(q,w)$ is a primary operator of dimension 1, the integrated one, ${\cal J}(q)$,
is a dimension zero operator and commutes with the Virasoro generators.
Therefore if a seed operator $V(k,z)$  is also a dimension 1 primary
operator,
the resultant bracket operator will be a dimension 1 primary operator.
Let $\mathcal{J}(q)$ and $\mathcal{V}(k)$ be the integrated operators,
\begin{align}
  \label{eq:41}
  \mathcal{J}(q)= \oint \frac{dw}{2\pi i}J(q,w) \,,
\qquad
\mathcal{V}(k)= \oint \frac{dz}{2\pi i}V(k,z) \,.
\end{align}
The state constructed by the action of $V^\text{br}$ is written, by
state--operator correspondence through the action of the commutator on the momentum vacuum, as $\commt{\mathcal{J}(q)}{\mathcal{V}(k)} \ket{0;0}$.
The physical state condition is
\begin{align}
0=&  \commt{L_n}{\commt{\mathcal{J}(q)}{\mathcal{V}}(k)} \ket{0;0}
=\left(
 \commt{\mathcal{J}(q)}{\commt{L_n}{\mathcal{V}}(k)}
- \commt{\mathcal{V}(k)}{\commt{L_n}{\mathcal{J}(q)}}
\right) \ket{0;0}
\,,\nn
\end{align}
for $n\geq 1$ where we have used Jacobi's identity.
$L_n$ represents a Virasoro generator.
Therefore it is easy to see that a sufficient condition for $V^\text{br}$
to be a physical vertex operator is both $J(q,w)$ and $V(k,z)$ being physical.
The question is what are necessary conditions. 
Let us take the bracket operator in $[\tilde{2}000]$ amplitude
as an example.
The bracket states corresponding to
$V^\text{br}_{(2)}(\widetilde{k},z)$ is
(here $k$ represents $k_1$ in the example)
\begin{align}
  \label{eq:45}
&  \bigg[
(\zeta_q \cdot k) (q \cdot \alpha_{-1}) (\zeta \cdot \alpha_{-1})
-(\zeta \cdot q) (q \cdot \alpha_{-1}) (\zeta_q \cdot \alpha_{-1})
+(\zeta_q \cdot \alpha_{-1})(\zeta \cdot\alpha_{-1})
\nn\\&\hskip1em
-(\zeta \cdot q)\zeta_q \cdot \alpha_{-2}
+\frac{1}{2}\left( (\zeta_q \cdot \zeta)-(\zeta_q \cdot k)(\zeta \cdot q) \right)
\left( (q \cdot \alpha_{-1})^2 + q\cdot \alpha_{-2} \right)
\bigg] \ket{0;\widetilde{k}}
\,,
\end{align}
and the physical state conditions
are 
\begin{align}
  \label{eq:46}
  0=& (\zeta \cdot q)(\zeta_q \cdot q)
\,,
\nn\\
0=&
 (\zeta_q \cdot q) \zeta_{\mu} 
+(\zeta \cdot k) \zeta_{q\mu}
+\big[(\zeta \cdot k)(\zeta_q \cdot k)-(\zeta \cdot q)(\zeta_q \cdot q)    \big]
q_\mu \,.
\end{align}
The first condition requires $\zeta \cdot q=0$ or $\zeta_q \cdot q=0$.
When $\zeta_q \cdot q=0$, the second condition says
$\zeta \cdot k=0$ or $\zeta_q^\mu = -(\zeta_q \cdot k)q^\mu$.
It is easy to see that when $\zeta_q \propto q$, the 
bracket state \eqref{eq:45}
identically vanishes.
So a sensible condition is $\zeta \cdot k=0$.
On the other hand, if we take $\zeta \cdot q=0$,
the second condition is 
\begin{align}
  \label{eq:47}
  (\zeta_q \cdot q) \zeta^\mu + (\zeta \cdot k) \zeta_q^\mu +(\zeta\cdot k)(\zeta_q \cdot k)q^\mu
=0 \,.
\end{align}
Contraction with $q_\mu$ leads to $(\zeta \cdot k)(\zeta_q \cdot q)=0$.
$\zeta_q \cdot q=0$ coincides with the previous choice,
while with $\zeta \cdot k=0$, \eqref{eq:47} implies $\zeta_q \cdot q=0$ unless 
$\zeta^\mu=0$ identically.
Therefore, the physical state conditions for the bracket state lead to 
the conditions, $\zeta \cdot k = \zeta_q \cdot q=0$, which are nothing but
the physical state conditions for each $J_q$ and $V(k)$.
So in this case, the sufficient conditions are also the necessary conditions.

However this may not be a general feature.
Indeed, in the case of $\alpha'q^2=-1$,
we find physical bracket states generated by a deformer operator with
an unphysical choice of polarizations.
The details of this example are presented
 in Appendix \ref{sec:bracket-operators}.
However, such physical states seem quite special, and in the following
discussion we confine ourselves in considering physical bracket states
generated by physical deformer and seed operators.

%%
%3.2
\subsection{Spectrum analysis of the bracket states}
\label{spectanlysis}

As seen, possible physical states obtained through the bracket operator
is governed by the physical polarizations for the deformer operator $J_q$
and the seed operator.

In the previous example, there are $25$ choices for each $\zeta_\mu$ and
$\zeta_{q\mu}$.
It is easy to see that the bracket state \eqref{eq:45}
are symmetric under the
exchange
of $\zeta$ and $\zeta_q$ when physical;
As seen, $\zeta_q=q$ makes \eqref{eq:45} trivially vanish,
while it is not difficult to check that $\zeta=k$ gives a null state.
Since $q^2=k^2=0$ and $k\cdot q=-1$, $k$ and $q$ are linearly independent,
which implies $\zeta \cdot q = \zeta_q \cdot k=0$ for physical states
of positive norm, and the statement follows.
Both $\zeta_\mu$ and $\zeta_{q\mu}$ are transverse to both
$k$ and $q$, and then
this choice of seed and deformer operators generates at most
$300$ physical states, while the total number of positive norm states
at level $2$ is $324$.

This counting will be more vividly illustrated by considering the
simplest case; namely,
both seed and deformer operators are tachyons,
$J_{(0)}(q,w)$ and $V_{(0)}(k,z)$
with $q^2=k^2=2$ ($\alpha'=1/2$).
The bracket operator $\{ \mathcal{J}_{(0)}(q) , V_{(0)}(k,z) \}$
is not trivial for $q\cdot k \leq -1$,
and a first few choices of $q \cdot k$ lead to
\begin{align}
  \label{eq:22}
& :e^{i \tilde{k} \cdot X}: \quad (q \cdot k =-1)
\,, \qquad
i\zeta_1 \cdot \partial X e^{i \tilde{k} \cdot X}
\quad ( q \cdot k =-2)
\,,\nn\\ & 
: \left[
-\zeta_{2\mu\nu}\partial X^\mu \partial X^\nu
+i\zeta_{2\mu} \partial^2 X^\mu
\right] e^{i\tilde{k} \cdot X} :
\quad ( q \cdot k = -3 )
\,, \cdots
\end{align}
where $\zeta_{1\mu} = q_\mu$,
$\zeta_{2\mu\nu}= q_\mu q_\nu /2$, and $\zeta_{2\mu}=q_\mu/2$.
These polarization tensors satisfy the physical state conditions and
then the bracket operators are physical.
In this case, we have only one state at each level.

In general, the number of physical states
of a given bracket operator is restricted by the numbers of
physical states of the seed and deformer operators and
is not much larger than
the product of these two numbers\footnote{%
When necessary conditions
agree with sufficient ones, the product gives an upper bound.
If not, there can be some extra physical states, 
but the number of them does not seem so large.
See Appendix~\ref{sec:bracket-operators}.}.
However, as seen from construction, in order to generate a bracket operator
at a given level,
there are infinitely many possible choices of seed and deformer
operators with $q \cdot k$ suitably chosen.
Therefore, missing physical states from a choice of seed
and deformer operators will be obtained from another choice.
These two different choices are in general
involved in different sets of exact relations.
Through a possible overlap of states, scattering amplitudes
are related to one another in a complicated way
and then are highly constrained.

We have observed that Moore's relation is quite powerful to relate
infinitely many scattering amplitudes in a very nontrivial way, and
these relations hopefully provide some trails of stringy symmetries.
As stated in the introduction, we will  carry out a first
concrete analysis by use of a couple of specific cases.
When we come to the consideration in massive inter-level relations,
there appear another complication in the choice of momenta and also physical
polarizations.
In the following sections, we consider the simplest choice
of the physical bracket operators
and physical states; namely, the ones from
physical seed and deformer operators, and the corresponding
states.
The more systematic analysis will be reserved for future study.

%4
\section{Kinematics of the four-point amplitudes}
\label{sec:4}

\subsection{Kinematic configuration} 

In this section, we shall give explicit solutions of the kinematic configuration both in the rest frame (of the first particle) and the center-of-momentum frame. 
All components of seed/bracket momenta can be expressed as functions
of the Mandelstam variables $s$ and $t$, and this will help us in
constructing various
polarization vectors needed for higher-spin amplitudes.

We start with the kinematic configuration of the
scattering processes in the rest frame of the first massive particle
($\widetilde{m}_1^2 \ne 0$).
For the sake of convenience, we take $\alpha' = 1/2$ in the following discussion. 
The following setup is the most economical ansatz which is compatible
with the momentum conservation.
\begin{align}
   q =&  (c_0 \hspace*{.1pc} , c_1 \hspace*{.1pc}, c_2 \hspace*{.1pc}, c_3 \hspace*{.1pc}, \vec{0}) \, ,
   \\
    \widetilde k_1  =   k_1 +q =&  ( \widetilde{k}_1^0 , \hspace*{.1pc} 0 \hspace*{.3pc}, \hspace*{.1pc} 0 \hspace*{.35pc}, \hspace*{.1pc} 0 \hspace*{.35pc}, \vec{0}) \,,
    \\
    k_2 =&  ( k_2^0 , k_2^1, \hspace*{.1pc} 0 \hspace*{.35pc},  \hspace*{.1pc} 0 \hspace*{.35pc},  \vec{0}) \,,
    \\
    k_3 =&  ( k_3^0 , k_3^1, k_3^2, \hspace*{.1pc} 0 \hspace*{.35pc},  \vec{0}) \,,
    \\
    k_4 =&  ( k_4^0 , k_4^1, k_4^2, \hspace*{.1pc} 0 \hspace*{.35pc},  \vec{0}) =  -\widetilde{k}_1-k_2-k_3 \,.
  \end{align}
In this rest-frame configuration, we can embed the seed and bracket momenta into a $(1 + 3)$-dimensional space-time, while the relevant physical momentum($\widetilde{k}_1, \, k_2 \sim k_4$) 
are confined within the  $(1 +2)$-dimensional scattering plane. 
Aside from the fourth momentum $k_4$ which is fixed by momentum
conservation, we have ten unknown components to be solved from the
five on-shell conditions
($\tilde{k}_1^2=-\tilde{m}_1^2$, $k_i=-m_i^2$ ($i=2,3,4$), and $q^2=m_q^2$)
and the three level number constraints
($\tilde{n}_1=\tilde{k}_1 \cdot q =n_1 - m_q^2$,
$n_2=k_2 \cdot q$, and $n_3= k_3 \cdot q$
($n_4$ condition is trivial due to the consistency condition
\eqref{eq:39}
when momentum conservation is satisfied)).
Hence it is natural to expect that we can solve all momenta in terms
of two Mandelstam variables, $s=-(\tilde{k}_1+k_2)^2$ and 
$t=-(\tilde{k}_1+k_3)^2$.
We also define $\tilde{s}=\tilde{k}_1 \cdot k_2$ and
$\tilde{t}=\tilde{k}_1 \cdot k_3$ for convenience. 

Through some algebraic manipulations, we find
\begin{align}
    \widetilde{k}_1^0 =&  \widetilde{m}_1 \,,
    \label{k0tilde}
    \\
     k_2^0 =& \frac{\widetilde{s}}{2\widetilde{m}_1}  \,,
\qquad
    k_2^1 = \frac{\delta_1 \sqrt{K_1(s)}}{2\widetilde{m}_1}  \,,
    \\
\label{eq:k3_comp}
    k_3^0 =&  \frac{\widetilde{t}}{2\widetilde{m}_1}\,,
    \quad
    k_3^1 = \frac{\delta_1   K_3(s,t)
           }{2 \widetilde{m}_1\sqrt{K_1(s)} } \,,
    \quad
    k_3^2 = \frac{\delta_2}{2\widetilde{m}_1}
                 \sqrt{\frac{K_1(s) \, K_2(t) -\big[K_3(s,t)]^2
                 }{K_1(s)}}
\,,
\end{align}
where we have defined
$K_1(s)=\widetilde{s}^2 -4\widetilde{m}_1^2m_2^2$,
$K_2(t)=\widetilde{t}^2-4\widetilde{m}_1^2 m_3^2$,
and
$K_3(s,t)=2 \widetilde{m}_1^2 \big(\widetilde{s}+\widetilde{t}+\widetilde{m}_1^2+m_2^2+m_3^2-m_4^2 \big)                +\widetilde{s}\widetilde{t}$
to make the equations brief.
$\delta_i=\pm 1$ ($i=1,2,3$) are introduced for the sign ambiguity.
On the other hand, the components of the bracket momenta $q$ are:
\begin{align}
     c_0 =&   - \frac{\widetilde{n}_1}{\widetilde{m}_1}  \,,
    \qquad
    c_1 =   \delta_1 \frac{2n_2 \widetilde{m}_1^2 -\widetilde{s} \widetilde{n}_1}{\widetilde{m}_1\sqrt{K_1(s)}} \,,
    \\
\label{c2}
    c_2 =&   \frac{\delta_2}{\widetilde{m}_1}
                 \frac{
K_1(s) (2\widetilde{m}_1^2 n_3 - \widetilde{t}\widetilde{n}_1)
                 -(2n_2 \widetilde{m}_1^2 -\widetilde{s} \widetilde{n}_1) K_3(s,t)
                 }{\sqrt{K_1(s) \big[
                 K_1(s) \, K_2(t)
                 -\big(K_3 (s,t) \big)^2
                 \big]}}  
\,,
    \\
    c_3 =&   \delta_3 \sqrt{-m_q^2+c_0^2-c_1^2-c_2^2} \,.
\label{c3} 
\end{align}
Note that
if we demand that all physical momenta have real components, then the items inside the square root should be positive. Hence,  we have the following inequalities,
\begin{align}
\hspace*{-2pc}
K_1(s) \geq 0 \,,
\qquad
K_1(s) \, K_2(t)
\geq 
\big(K_3(s,t) \big)^2 \,,
\qquad
c_0^2 \geq   m_q^2 + c_1^2+ c_2^2 \,.
\end{align}

Having obtained the expressions of various momenta in the rest frame we can derive the kinematic configuration in the center-of-momentum (CM)
frame, $\widetilde{k}_1^{1'} + k_2^{1'} = 0$, by boosting along $x^1$
direction
with velocity $\beta = K_1(s)/(\tilde{s}+2\tilde{m}_1^2)$.
For $\delta_{1,2,3}=1$ choice, we find (assuming  $s>0$)
\begin{align}
\label{eq:k1t_CM}
 \tilde  k^{(CM)}_1 =&
\frac{1}{2\sqrt{s}} \left(
s+\tilde{m}_1^2-m_2^2 , -\sqrt{K_1(s)} , 0, 0, \vec{0} 
\right) \,,
\\
\label{eq:k2_CM}
 k^{(CM)}_2 =&
\frac{1}{2\sqrt{s}} \left(
s-\tilde{m}_1^2+m_2^2 , \sqrt{K_1(s)} , 0, 0, \vec{0} 
\right) \,,
\\
\label{eq:k3_CM}
 k^{(CM)}_3 =&
 \left(
- \frac{\tilde{s}+\tilde{m}_1^2+m_2^2+m_3^2-m_4^2 }{2\sqrt{s}}, 
k_3^{(CM)1}, k_3^2 \, 0 ,\vec{0}
\right) \,,
\\
\label{eq:q_CM}
q^{(CM)} =&
\left(
-\frac{\tilde{n}_1+n_2}{\sqrt{s}} ,
 \frac{-\tilde{s}(\tilde{n}_1-n_2)-2m_2^2\tilde{n}_1 +2\tilde{m}_1^2 n_2}{\sqrt{s}\sqrt{K_1(s)}}
, c_2 ,c_3 , \vec{0}
\right)
\end{align}
where
\begin{align}
  \label{eq:31}
  k_3^{(CM)1} =&
\frac{1}{2\sqrt{s}\sqrt{K_1(s)}}
\big[
\tilde{s}^2 + 2\tilde{s}\tilde{t} + \tilde{s}(3\tilde{m}_1^2+m_2^2+m_3^2-m_4^2) + 2\tilde{t}(\tilde{m}_1^2 + m_2^2)
\nn\\&\hskip8em
 +2\tilde{m}_1^2(\tilde{m}_1^2+m_2^2+m_3^2-m_4^2) \big] \,,
\end{align}
and $k_3^2$, $c_2$, and $c_3$ are the same as the rest frame configuration,
\eqref{eq:k3_comp}, \eqref{c2} and \eqref{c3}.

Our main interest is to examine the relations between high-energy symmetry \` a la Gross and the exact identities as derived from bracket algebra. 
In the case of fixed-angle high-energy scattering, we take $s,t \rightarrow \infty$, and keep $t/s$ fixed, and consequently,
\begin{equation}
   \widetilde{s} = s + \mathcal{O}(1)\, ,
\qquad
\widetilde{t} = t + \mathcal{O}(1) = -\dfrac{s}{2}(1 - \cos \theta_{CM} ) + \mathcal{O}(1) \, ,
\end{equation}
where $\theta_{CM}$ is the scattering angle in the CM frame. 
The leading-order expressions are given by
(with the sign factors $\delta_i$ restored)
\begin{align}
  \tilde k_1^{(CM)} =& \frac{\sqrt{s}}{2} (1,-\delta_1 , 0,0, \vec{0}) \,,
  \label{eq:gen_CM_mom_HE1}
\qquad
k_2^{(CM)} = \frac{\sqrt{s}}{2} (1,\delta_1, 0,0, \vec{0}) \,,
%  \label{eq:gen_CM_mom_HE2}
\\
k_3^{(CM)} =& \frac{\sqrt{s}}{2} \left( -1, \delta_1 \cos\theta_{CM} ,
\delta_2 \sqrt{1-\cos^2\theta_{CM}} , 0 , \vec{0}
\right) \,,
  \label{eq:gen_CM_mom_HE3}
\\
q^{(CM)}=&
\frac{-1}{\sqrt{s}} \left(
\tilde{n}_1+n_2 , \,
\delta_1( \tilde{n}_1-n_2) , \,
 -\delta_2  \frac{(n_2-\tilde{n}_1)(1-\cos\theta_{CM})+2n_3+2\tilde{n}_1 }{\sqrt{1-\cos^2\theta_{CM}}} ,
 \right.\nn\\&\left.\hskip3em
 -\delta_3 \sqrt{s} \sqrt{-m_q^2} , \vec{0}
 \right) \,.
   \label{eq:gen_CM_mom_HEq}
\end{align}
One can see that
$\delta_1$ and $\delta_2$ are responsible for covering all the kinematic
range by use of this parametrization, while $\delta_3$
has no physical importance.
Note that the third spatial component of the momentum $q$ becomes pure imaginary in this limit, when $m_q^2>0$.
However, all physical momenta are real.

\subsection{Complex momenta and Lorentz transformations}

As seen in \eqref{eq:gen_CM_mom_HE1}--\eqref{eq:gen_CM_mom_HEq},
in the high-energy limit $s \rightarrow \infty$,
the momenta $\tilde{k}_1$, $k_2$, $k_3$, and $k_4$ posses real components
for generic choices of masses and $q \cdot k_i$, while
$q$ will develop a complex component when it corresponds to a massive state.
One uses $\tilde{k}_1$, $k_2$, $k_3$, and $k_4$ as the momenta for external
particles to calculate a scattering amplitudes, and the amplitude
is regarded as a physical scattering process in
a high energy regime.
In the calculation of scattering amplitudes,
$q$ appears only through polarization tensors for bracket states.

Moore's prescription relates a set of scattering amplitudes 
in which different operators are deformed,
and each amplitude carries different sets of momenta.
For examples discussed in the paper,
 one of them has 
$\tilde{k}_1=k_1+q$, ${k}_2=k_2$, $k_3$, and $k_4$
and the other 
$k_1$, $\tilde{k}_2=k_2+q$, $k_3$, and $k_4$.
Thus, $k_1$ and $\tilde{k}_2$ may become complex in the high-energy limit.
In view of Moore's relation as an identity among analytic functions of
momenta and polarization invariants, it is not a problem.
However, one may be worried about whether the relation is understood
as a relation
among \textit{physical} amplitudes, at least in an asymptotic regime
of our main interest.
The latter momentum set is characterized by
the masses $m_1, \tilde{m}_2, m_3, m_4, m_q$ and the integers $n_1,
\tilde{n}_2, n_3, n_4$. Since the general formulas
\eqref{eq:k1t_CM}--\eqref{eq:q_CM} defines real external momenta
for the given set of the parameters, we may work with these momenta to
compute physical scattering amplitudes.
Since the amplitude is a function of Lorentz invariants, such as
$k_i \cdot k_j$ or $\zeta \cdot k_i$,
it defines an
equivalent amplitude if the invariants are the same.
This condition is satisfied if there is a ``Lorentz transformation''
$SO(1,25; \bC)$ (or $SO(1,3; \bC)$ in practice)
that relates these two configurations.
For the kinematic configuration in the
case of $\mathcal{A}[\tilde{3}000]=\mathcal{A}[1\tilde{2}00]$,
we have shown it by explicitly constructing the transformation matrix,
and we conclude that the exact symmetry identities indeed relate
various physical scattering amplitudes with real momenta.

\subsection{Scattering and and $q$-orthonormal helicity bases for polarizations}

We first present a general discussion of constructing a new orthonormal basis which
is suitable for the study of Moore's relation, based on the helicity representation
with respect to the first particle.
Assume that $\tilde{k}_1$ is a momentum for a massive state.
Let $e^P$, $e^L$, $e^T$, $e^I$, and $e^{J_i}$ be the helicity vectors with respect to
$\tilde{k}_1$
in the CM frame.
Here, $e^P \propto \tilde{k}_1$ is the momentum direction of the first
particle. $e^L$ is the longitudinal vector, namely a unit vector
parallel to $\tilde{k}_1$ on the scattering plane.
$e^T$ is the transverse vector lying on the scattering plane.
$e^I$ is one of the other transverse vectors which has an overlap with
$q$. $e^{J_i}$ ($i=1,\cdots,22$) are the rest of the transverse vectors,
chosen so that they are orthogonal to all the momenta in question.
The completely transverse vectors $e^{J_i}$ are not relevant for the
discussion here, and we neglect them for the time being.

Since these vectors serve a natural basis for polarization tensors when we
discuss scattering amplitudes in the CM frame, we call them the
scattering helicity basis.
They are also convenient basis to
analyze the physical state conditions for the massive first
particle\cite{Chan:2005qd}.
However, when the first particle corresponds to a bracket operator as
for our examples, the physical state conditions are more neatly
written down by use of a basis regarding the deformation momentum $q$
as we shall see.
% See Fig.~\ref{fig:eT-eI} and \eqref{eq:eP_2000}--\eqref{eq:eT-eI_2000} for illustration.
$\tilde{k}_1$ and $q$ have the following expression on the scattering
helicity basis,
\begin{align}
  \tilde{k}_1 =& \sqrt{\tilde{m}_1^2} e^P \,,
\qquad
q= c_P e^P + c_L e^L + c_T e^T + c_I e^I \,,
\end{align}
where $c_P$ is determined by the condition $q \cdot \tilde{k}_1 =\tilde{n}_1$
as $c_P=-\tilde{n}_1/\sqrt{\tilde{m}_1^2}$, while the other coefficients depend on the choice of $k_2$, $k_3$, and $k_4$, as we have just seen.
We are now going to define a new set of orthonormal basis vectors with which
$q$ takes the following simple form, $q= c_P e^P + c_Q e^Q$, 
with $e^Q$ being a unit vector defined simply by
$c_Q e^Q = c_L e^L + c_T e^T + c_I e^I$
and $c_Q^2 =\sqrt{c_L^2+c_T^2+c_I^2}= c_P^2 - m_q^2$.
In the subspace spanned by $e^L$, $e^T$ and $e^I$, we define another
two unit vectors orthogonal to $e^Q$; we choose $e^{T_q}$ to be
purely spatial and $e^{I_q}$ is the orthogonal complement to $e^Q$ and
$e^{T_q}$ in this subspace.
The sign ambiguity has no physical importance.
$e^{T_q}$, $e^{I_q}$, and $e^Q$, together with $e^P$, form a new orthonormal basis
which we call the $q$-orthonormal basis.
On this basis, $k_1$ and $q$ are represented as
\begin{align}
  \label{eq:19}
  {k}_1 =& \left( \sqrt{\tilde{m}_1^2} -c_P \right) e^P
-c_Q e^Q
 \,,
\qquad
q= c_P e^P + c_Q e^Q \,,
\end{align}
and then the physical state conditions for the seed operator $V(k_1,z)$
and the deformer $J(q,z)$ are written down by use of
``longitudinal-like'' $e^Q$
and ``transverse''
unit vectors $e^{T_q}$ and $e^{I_q}$ (and also completely transverse
vectors $e^{J_i}$).
In this basis, $e^P$ is not the momentum direction of $q$ and then $e^Q$
is not really longitudinal.
However, it turns out to be convenient to keep $e^P$ dependence
explicitly, since $e^P$ is directly related to decoupling states from 
 bracket operators.

In summary, we have defined a new set of the unit orthogonal vectors
\begin{align}
\label{eq:transf_formura}
  e^{A'} =& \sum_{a'=L,T,I} C^{A'}{}_{a'} e^{a'} \,,
\end{align}
where $A'=T_q, I_q, Q$ and
the explicit form of the
transformation matrix $C^{A'}_{a'}$ is
\begin{align}
\label{eq:C_Tq}
  C^{T_q}{}_L=& 0 \,,
\qquad
C^{T_q}{}_T= \frac{c_I}{\sqrt{c_T^2+c_I^2}} \,,
\qquad
C^{T_q}{}_I = \frac{-c_T}{\sqrt{c_T^2+c_I^2}} \,,
\\
\label{eq:C_Lq}
  C^{I_q}{}_L=& \frac{\sqrt{c_T^2+c_I^2}}{c_Q} \,,
\quad
C^{I_q}{}_T= \frac{-c_L c_T}{c_Q\sqrt{c_T^2+c_I^2}} \,,
\quad
C^{I_q}{}_I =\frac{-c_L c_I}{c_Q\sqrt{c_T^2+c_I^2}} \,,
\\
\label{eq:C_Q}
  C^{Q}{}_L=& \frac{c_{L}}{c_Q} \,,
\qquad
C^{Q}{}_T= \frac{c_T}{c_Q} \,,
\qquad
C^{Q}{}_I =\frac{c_I}{c_Q} \,.
\end{align}
Since the both $(e^L,e^T,e^I)$ and $(e^{I_q}, e^{T_q}, e^Q)$ are
orthonormal with the positive metric, the transformation matrix is orthogonal,
namely
\begin{align}
  \sum_{a'=L,T,I} C^{A'}{}_{a'}  C^{B'}{}_{a'}  = \delta^{A' B'} \,,
\qquad
  \sum_{A'=I_q,T_q,Q} C^{A'}{}_{a'}  C^{A'}{}_{b'}  = \delta_{a' b'} \,,
\end{align}
where $A', B'= I_q, T_q, Q$ and $a',b'=L,T,I$.
In later sections we shall see the advantage of this
 new basis to represent usual scattering amplitudes
in the CM frame and
and also DDF amplitudes.

\subsection{Kinematics for Case studies}
\label{sec:kinem-case-stud}

Based on the general discussion so far, we write down the explicit
kinematic configurations for the examine we examine in this paper,
for reference.

\paragraph{Case study I: $\mathcal{A}[\tilde{2}000]=\mathcal{A}[1\tilde{1}00]$}

In this case,
the scattering helicity basis with respect to $\tilde{k}_1$
($\tilde{m}_1^2=2$) is given by
\begin{align}
  e^P=& \frac{1}{2\sqrt{2s}} \big( s+4 , -\sqrt{s^2+16} , 0,0 \big)
\,,
\label{eq:eP_2000}
\\
e^L=& \frac{1}{2\sqrt{2s}} \big( \sqrt{s^2+16} , -(s+4), 0,0 \big)
\,,
\label{eq:eL_2000}
\\
e^T=& (0,0,1,0) \,,
\qquad e^I =(0,0,0,1) \,.
\label{eq:eT-eI_2000}
\end{align}
The momenta in $\mathcal{A}[\tilde{2}000]$ side are
obtained by the general formulas \eqref{eq:k1t_CM}--\eqref{eq:q_CM}
and the auxiliary vector $q$ is represented in this basis as
\begin{align}
q=&
 \frac{1}{\sqrt{2}}e^P
-\frac{s-4}{\sqrt{2}\sqrt{s^2+16}} e^L
+\frac{4(s+2t+4)}{\sqrt{s^2+16}\sqrt{f_1(s,t)}} e^T
+\frac{2\sqrt{f_2(s,t)}}{\sqrt{f_1(s,t)}}  e^I
\,,
\end{align}
where $f_1(s,t)=32-st(s+t+4)$
and $f_2(s,t)=-4-t(s+t+4)$.
The basis vectors of the $q$-orthonormal basis
are obtained
from the transformation formulas
\eqref{eq:transf_formura}--\eqref{eq:C_Q}
in the previous subsection.

For the right hand side (RHS), ${\cal A}[1\tilde{1}00]$,
we need to prepare another scattering helicity basis with
respect to $k_1$ and $\tilde{k}_2=k_2+q$.
Let $e^{P_1} , e^{P_2}, e^{T_R}, e^{I_R}$, and $e^{J_i}$ be
basis vectors in question.
$e^{P_1}$ and $e^{P_2}$ are momentum polarization with respect to
$k_1$ and $\tilde{k}_2$ respectively.
Since they are null, there are no $L$-directions.
$e^{T_R}$ is on the RHS scattering plane (now spanned by $\vec{k_1}$ and
$\vec{k}_3$) and orthogonal to $k_1$.
$e^{I_R}$ is a unit vector perpendicular to the RHS scattering plane.
The purely transverse directions $e^{J_i}$ are common on both hands
sides, and we use the same basis vectors.
By use of $q$-orthonormal basis, they are represented as
\begin{align}
e^{P_1}=& \frac{1}{\sqrt{2}}e^P - \frac{1}{\sqrt{2}}e^Q \,,
\qquad
e^{P_2}= \frac{s+2}{2\sqrt{2}} e^P + \frac{s-2}{2\sqrt{2}}e^Q
-\sqrt{s} e^{I_q} \,,
\\
e^{T_R}=& e^{T_q} \,,
\qquad \qquad
  e^{I_R}=
-\frac{\sqrt{2}}{\sqrt{s}} e^P
+\frac{\sqrt{2}}{\sqrt{s}} e^Q
+e^{I_q} 
 \,.
\end{align}
It is straightforward to check that $e^{I_R}$ is orthogonal to
$e^{P_1}$ and $e^{P_2}$.

\paragraph{Case study I{}I: $\mathcal{A}[\tilde{3}000]=\mathcal{A}[1\tilde{2}00]$}

The momenta for the case are obtained from the general formulas
\eqref{eq:k1t_CM}--\eqref{eq:q_CM}.
The helicity basis with respect to $\tilde{k}_1$ is
\begin{align}
  \label{eq:1}
    e^P=& \frac{1}{\sqrt{\tilde{m}_1^2}}\tilde{k}_1
= \frac{1}{4\sqrt{s}} \left( s+6, \, -\sqrt{F_3(s)} , 0,0 \right)
\,,\\
  e^L=&
\frac{1}{4\sqrt{s}} \left(\sqrt{F_3(s)}, \, -(s+6),   0,0 \right)
\,,\\
  e^T=& (0,0,1,0)
\,,\qquad
  e^I= (0,0,0,1)
\,,
\end{align}
and the coefficients of $q$ in this basis are obtained by $c_A=\pm e^A_\mu
q^\mu$ for $A=P,L,T,I$ (the negative sign is for $e^P$).
Through $c_A$ one can generate $q$-orthonormal basis easily.
The helicity basis with respect to $k_1$
 for $\mathcal{A}[1\tilde{2}00]$ side
is constructed 
in a similar way to the previous example, in the $q$-orthonormal
basis, as
\begin{align}
e^{P_2}=&
 \frac{1}{\sqrt{2}} \tilde{k}_2
=-\frac{\sqrt{2}(s+4)}{8} e^P - \frac{3\sqrt{2}(s-4)}{8} e^Q
+ \frac{\sqrt{F_2(s)}}{2}e^{I_q} 
\,,\nn\\
e^{L_2}=&
\frac{-\sqrt{2}(s^2+2s-16)}{8(s-2)} e^P
+\frac{\sqrt{2}(3s^2-18s+32)}{8(s-2)} e^Q
- \frac{\sqrt{F_2(s)}}{2} e^{I_q}
\,,\nn\\
  e^{T_R}=& 
\frac{-(s-4)(s+2t+2)}{2(s-2)\sqrt{F_5(s,t)}} e^P
+\frac{F_6(s,t)}{2(s-2)\sqrt{F_5(s,t)}} e^Q
+\frac{F_6(s,t)}{\sqrt{2 F_2(s) F_5(s,t)}} e^{I_q}
\nn\\&
+\frac{-F_7(s,t)\sqrt{F_1(s,t)}}{F_4(s,t)\sqrt{2F_2(s)F_5(s,t)}} e^{T_q}
\,,\nn
\\
e^{I_R}=&
\frac{\sqrt{F_1(s,t)}}{2\sqrt{F_5(s,t)}} \big( e^P -e^Q \big)
+\frac{F_6(s,t)}{\sqrt{2 F_2(s) F_5(s,t)}} e^{T_q}
+\frac{F_7(s,t)\sqrt{F_1(s,t)}}{F_4(s,t)
  \sqrt{2F_2(s) F_5(s,t)}} e^{I_q}
\,,\nn
\end{align}
where $  e^{P_1}= k_1 = 2e^P -q$.
$e^{L_2}$ is the longitudinal unit vector for $\tilde{k}_2$,
and
\begin{align}
%  \label{eq:9}
    F_1(s,t)=&
    2\,{s}^{2}t-{s}^{2}-12\,st+2\,s{t}^{2}+16\,s-96-16\,{t}^{2}-32\,t 
\,,\nn\\
F_2(s)=& -{s}^{2}+10\,s-20 \,,
\qquad
F_3(s)= s^2-4s+36 \,,\nn
\\
F_4(s,t)=& 72-st(s+t+2) 
\,,\qquad
F_5(s,t)= -st(s+t+2)-8s+8 
\,,\nn
\\
F_6(s,t)=&
{s}^{2}-2\,s+2\,st-8-8\,t
\,,\nn\\
F_7(s,t)=&{s}^{3}t+{s}^{2}{t}^{2}-2\,s{t}^{2}-4\,st-72\,s+144
\,.\nn
\end{align}

%5
\section{Exact identities in various bases}
\label{sec:change_basis}

In this section, we discuss the relation of physical bracket
states with standard positive norm/DDF states.
As seen in Section \ref{confinv}, 
a bracket operator is physical when a deformer and a seed operators on
which the bracket operator is based are physical, and physical state
conditions for the deformer and seed operators are simply solved by
use of the $q$-orthonormal basis introduced in Section \ref{sec:4}.
As we will see, Moore's relation takes a simple form
when it is represented in terms of
amplitudes with $q$-orthonormal polarizations. Especially,
the coefficients in the relations are found to be $t$-independent.
However, the $q$-orthonormal basis is constructed with respect to a
deformation momentum $q$ which does not show up in physical momenta
for scattering amplitudes
and then it does not respect certain physical symmetries.
We therefore want to represent exact relations in a physical basis
--- usual helicity basis with respect to a momentum for an external particle.
Further transformation to a DDF basis is advantageous when we discuss
high-energy symmetries as explained in Section
\ref{sec:high-energy-linear}.

The translation involves energy dependent transformation coefficients
connecting different bases and the expressions of the coefficients are
fixed by bracket operation and the choice of $q$. As we will see, the
high-energy expansions of these coefficients provide proportional
constants of high-energy linear relations.

\subsection{Exact identities for Case study I}
\label{sec:bracket-states-terms-2000-1100}

In this subsection, we discuss the bracket states that appear in
${\cal A}[\tilde{2}000]={\cal A}[1\tilde{1}00]$ relation
and their decompositions into scattering helicity bases.

Since we are interested in scattering amplitudes for physical processes,
we need to impose physical state conditions on bracket states.
As discussed in Section \ref{sec:3}, 
the bracket operator is automatically physical
for a physical choice of
the seed polarization $\zeta_1$ and the
deformer polarization $\zeta_q$.
Physical conditions for the deformer and the seed operators
are met easily by use of $q$-orthonormal basis
introduced in Section \ref{sec:4} as
\begin{align}
  \zeta_1=e^A , \quad \zeta_q =& e^B \,,
\qquad A, B= {T_q}, I_q, {J_i} \,.
\label{eq:25}
\end{align}
By plugging \eqref{eq:25} into
\eqref{eq:26}, \eqref{eq:27}, and \eqref{eq:28},
the polarization tensors of the bracket states are written as
\begin{align}
  \zeta^{(2)}_{\mu\nu}(\zeta_1, \zeta_q) =&
e^A_{(\mu} e^B_{\nu)} + \frac{\delta^{AB}}{2} q_\mu q_\nu \,,
\nn\qquad
\zeta^{(2)}_\mu(\zeta_1,\zeta_q)= \frac{\delta^{AB}}{2}q_\mu \,,
\nn\\
 \zeta_{R\mu}(\zeta_q) =& e^B_\mu + k_2^B q_\mu\,.
\label{eq:br_hel_2000-1100}
\end{align}
The two polarization tensors on the first line are for the level 2 bracket
operator that appears in ${\cal A}[\tilde{2}000]$ calculation, while
the last one is for the level 1 bracket operator on ${\cal
  A}[1\tilde{1}00]$ side.
We may write the bracket operators of these choices of the
polarization tensors as $V_{(2)}^{\text{br}\, AB}(\tilde{k}_1,x)$
and $V_{(1)}^{\text{br}\, B}(\tilde{k}_2,1)$.
We are interested in the four-point amplitudes with these operators
inserted,
\begin{align}
 \mathcal{T}_{\text{br}[\tilde{2}000]}^{AB} =&
\int_0^1 dx 
\vev{V_{(2)}^{\text{br}\, AB}(\tilde{k}_1,x)
V_{(0)} (k_2, 0) V_{(0)} (k_3,1)V_{(0)} (k_4, \infty)}
\,,\\
 \mathcal{T}_{\text{br}[1\tilde{1}00]}^{A|B} =&
\int_0^1 dx 
\vev{V_{(1)}^A(k_1,x)
V_{(1)}^{\text{br}\, B}(\tilde{k}_2,1)
V_{(0)} (k_3,1)V_{(0)} (k_4, \infty)}
\,,
\end{align}
where the vertical line in the superscript of the second amplitude
denotes the separation of the first and the second particles.
In terms of these ``bracket amplitudes,'' Moore's relation
trivially reads
\begin{align}
  \label{eq:br_rel_2000-1100}
  \mathcal{T}_{\text{br}[\tilde{2}000]}^{AB} 
=& \mathcal{T}_{\text{br}[1\tilde{1}00]}^{A|B} 
\,.
\end{align}
Since the bracket operators are
vertex operators with specific forms of the polarization
tensors,
we may write Moore's relation in terms of the amplitudes
associated with conventional scattering amplitudes such as
\begin{align}
\int_0^1 dx
%(\text{ghost})
&
 \vev{ V_{(2)}(\tilde{k}_1,x;\zeta)
\,  V_{(0)}(k_2,0)
\,  V_{(0)}(k_3,1)
\,  V_{(0)}(k_4,\infty) }
\nn\\=&
F_{s-t}
\left[
\zeta_{\mu\nu} \mathcal{T}_{[2000]}^{\mu\nu} 
+ \zeta_\mu \mathcal{T}_{[2000]}^\mu
\right]
\,,
\label{eq:2000_sc_general}
\end{align}
where
\begin{align}
  F_{s-t} =& \frac{\Gamma \big(-\alpha s'-1)\Gamma \big(-\alpha' t-1) }{\Gamma(\alpha' u +2 )} \,,
\label{F_s-t}
\end{align}
is ``Veneziano-like'' part which is responsible for the soft behavior
in the high-energy regime.
We take out this factor from scattering amplitudes as a common factor,
and then $\mathcal{T}_{[2000]}^{\mu\nu}$ and
$\mathcal{T}_{[2000]}^{\mu}$
are the rest of ``polynomial'' part.
In the same manner we define $\mathcal{T}_{[1100]}^{\mu | \nu}$,
and Moore's relation is written in the following form,
\begin{align}
  {\cal T}_{[2000]}^{AB}
+\frac{\delta^{AB}}{2} \left(
{\cal T}_{[2000]}^{qq} + {\cal T}_{[2000]}^{q} 
\right)
=
{\cal T}_{[1100]}^{A|B}
+ k_2^B {\cal T}_{[1100]}^{A|q} \,,
\end{align}
where ${\cal T}_{[2000]}^{AB}=e^A_\mu e^B_\nu {\cal
  T}_{[2000]}^{\mu\nu}$, ${\cal T}_{[2000]}^{q}=q_\mu {\cal
  T}_{[2000]}^{\mu}$,
and so on.
Since $q=\frac{1}{\sqrt{2}}\big( e^P + e^Q \big)$,
the second terms of the both sides contain $e^P$ components.
The amplitudes with this component are related to vanishing amplitudes
due to the decoupling of zero norm states.
After dropping such trivial part, we can write the both hands
sides in terms of the $q$-transverse polarizations as
\begin{align}
&
{\cal T}^{AB}_{[2000]} -\frac{\delta^{AB}}{20}
\left(
-4 {\cal T}^{{Q}{Q}}_{[2000]}
+{\cal T}^{I_q I_q}_{[2000]}
+{\cal T}^{T_q T_q}_{[2000]}
+\sum_{i=1}^{22} {\cal T}^{J_i J_i}_{[2000]}
\right)
\nn\\=&  
{\cal T}^{A|B}_{[1100]} + \frac{2 k_2^B}{s+2}
\left(
\sqrt{s} {\cal T}^{A|I_q}_{[1100]}
+\sqrt{2} {\cal T}^{A|{Q}}_{[1100]}
\right)
\,,
\end{align}
where $A,B= T_q , I_q, J_i$.
This equality is exact and holds for arbitrary $s$ and $t$.
The coefficients
in the equality are almost just constants
and even non-constant coefficients are simple functions of $s$,
since
$k_2^{I_q}= -\sqrt{s}$,
$k_2^{{Q}} = \frac{s-4}{2\sqrt{2}}$, and
$k_2^{T_q}=
k_2^{J_i}=0$.
Especially, the coefficients are $t$ (therefore the scattering angle)
independent.
There appear five independent relations with respect to the
choice of $A$ and $B$.
This is first our observation; the exact identity relation takes
a particularly simple form with projection onto $q$-orthonormal basis.
Hence, the deformation momentum $q$ also provides a natural frame to describe
the exact identity.

However, when we look at the relation as a relation among physical
scattering amplitudes,
$q$ does not explicitly appear as a momentum of external particles
but is implicitly encoded in a specific form of deformed polarization
tensors.
Therefore transverse projections with respect to $q$ does not have
manifest physical significance.
We thus rewrite the relation in terms of amplitudes in scattering
helicity basis which is standard basis to describe scattering amplitudes
in the CM frame.

\paragraph{Level 2 bracket state in $\mathcal{A}[\tilde{2}000]$ amplitude}
\label{sec:level-2-bracket-2}

We first write bracket states in terms of scattering helicity states,
with zero norm states dropped.
The level 2 bracket state corresponding to
$V^\text{br}_{(2)}(\tilde{k}_1,z; \zeta^{(2)})$ is
rewritten as\footnote{%
$\ket{0;\tilde{k}_1}$ is the tachyon state with momentum $\tilde{k}_1$.}
\begin{align}
&
\ket{\zeta_1=e^A, \zeta_q=e^B; \tilde{k}_1}_\text{br}
=
\bigg[ \alpha_{-1}^{AB} + \frac{\delta^{AB}}{2} \big(\alpha_{-1}^{qq} + \alpha_{-2}^q \big)
\bigg] \ket{0;\tilde{k}_1}
\nn\\=&
\bigg[
\left( C^{(A}{}_{a}C^{B)}{}_{b} 
+\frac{\delta^{AB}}{4}C^Q{}_a C^Q{}_b
\right)\alpha_{-1}^{ab}
%\nn\\&\hskip1em
+2 
\left( C^{(A}{}_a C^{B)}{}_L
+\frac{\delta^{AB}}{4}C^Q{}_a C^Q{}_L
\right)
 \alpha_{-1}^{aL}
\nn\\&\hskip1em
+\left( C^{A}{}_{L}C^{B}{}_{L} 
+\frac{\delta^{AB}}{4}C^Q{}_L C^Q{}_L
\right)\alpha_{-1}^{LL}
\nn\\&\hskip1em
+\frac{\sqrt{2}C^Q{}_a \delta^{AB}}{4} \left(
\alpha_{-2}^a + \sqrt{2}\alpha_{-1}^{aP}
\right)
%\nn\\&\hskip1em
+\frac{\sqrt{2}C^Q{}_L \delta^{AB}}{4} \left(
\alpha_{-2}^L + \sqrt{2}\alpha_{-1}^{LP}
\right)
\nn\\&\hskip1em
+\frac{\delta^{AB}}{20} \left(
5\alpha_{-1}^{PP} + 5\sqrt{2} \alpha_{-2}^P \right)
\bigg] \ket{0;\tilde{k}_1}
\,,
\label{eq:br_2000}
\end{align}
where $A,B=T_q, I_q, J_i$ and $a,b=T,I,J_i$.
We summarize a product of oscillators of the same level
as $\alpha_{-1}^{ab} \equiv \alpha_{-1}^a \alpha_{-1}^b$, and
$\alpha_{-1}^q = q \cdot \alpha_{-1}$.
${C^A}_a$ is the transformation matrix defined in (\ref{eq:transf_formura}).
In the second from the last line,
these two specific linear combinations are proportional
to zero norm states of this level.
The explicit forms of zero norm states are summarized in Appendix
\ref{sec:from-covar-posit}.
In the last line, we consider the difference with one of the zero norm state
as
\begin{align}
% \frac{\delta^{AB}}{20}
 \left(
5\alpha_{-1}^{PP} + 5\sqrt{2} \alpha_{-2}^P \right)
 \ket{0;\tilde{k}_1}
=&
% -\frac{\delta^{AB}}{20}
- \bigg(
\alpha_{-1}^{LL} + \sum_{a} \alpha_{-1}^{aa} 
\bigg) \ket{0;\tilde{k}_1}
%\nn\\&
+\ket{ZN_1; \tilde{k}_1} \,.\nn
\end{align}
By dropping all the zero norm state parts,
we find that the bracket state \eqref{eq:br_2000} can be written
in terms of positive norm states, up to zero norm states, as
\begin{align}
  \eqref{eq:br_2000} =&
\bigg(
\sum_{a',b'} G^{AB}_{a'b'} \alpha_{-1}^{a'b'}
+G^{AB} \sum_{a'}  \alpha_{-1}^{a'a'} 
\bigg) \ket{0;\tilde{k}_1}
%+(\text{zero norm states})
 \,,
\end{align}
where
$a',b'=L,T,I,J_i$, namely the longitudinal direction and
the transverse directions.
The coefficients are
\begin{align}
\label{eq:18}
G^{AB}_{a'b'}=&
\left( C^{A} C^{B} 
+\frac{\delta^{AB}}{4}C^Q C^Q
\right)_{(a'b')} 
\,,
\qquad
 G^{AB}=
 -\frac{\delta^{AB}}{20}
 \,.
\end{align}
$(C^AC^B)_{(a'b')}$ 
is the symmetrization,
$(C^AC^B)_{(a'b')} \equiv \frac{1}{2} \big( C^A_{a'} C^B_{b'} + C^A_{b'} C^B_{a'} \big)$.

\paragraph{Level 1 bracket state in $\mathcal{A}[1\tilde{1}00]$ amplitude}
\label{sec:level-1-bracket}

We now move on to the $\tilde{m}_2^2=0$ bracket state.
The state is fairly simple,
\begin{align}
  \ket{\zeta_q=e^B; \tilde{k}_2}_\text{br} =&
\bigg( \alpha_{-1}^B + k_2^B \alpha_{-1}^q
\bigg) \ket{0;\tilde{k}_2} \,.
\label{eq:br_lv1}
\end{align}
Note that $e^B$ is transverse to $q$ but not to $k_2$.
We have chosen the polarization tensor of the deformer $\zeta_{q\mu}$
so that the polarization tensor associated with
this bracket state satisfies the transversality condition
with respect to $\tilde{k}_2$,
\begin{align}
\tilde{k}_2 \cdot \zeta_{R}=0 \,,
\qquad  \zeta_{R\mu}(\zeta_q=e^B) = e^B_\mu + k_2^B q_\mu 
\,.
\end{align}
Therefore, this bracket states can be rewritten as a linear combination
of the momentum and the transverse oscillators,
\begin{align}
  \eqref{eq:br_lv1} =&
\bigg( \tilde G^B_{P_2} \alpha_{-1}^{P_2} 
+ \sum_{\tilde{a}} \tilde G^B_{\tilde{a}} \alpha_{-1}^{\tilde{a}}
\bigg) \ket{0;\tilde{k}_2} \,,
\end{align}
where $\tilde{a}=T_R, I_R, J_i$,
the transverse directions with respect to $k_1$ and $\tilde{k}_2$,
and the coefficients are determined to be
\begin{align}
  \tilde G^B_{P_2} =& - e^{P_2} \cdot \zeta_{R} \,,
\qquad
\tilde G^B_{\tilde{a}} = e^{\tilde{a}} \cdot \zeta_{R} \,.
\end{align}
Here, we put the tilde for coefficients on the right-hand side
(namely $\mathcal{A}[1\tilde{1}00]$) states for distinction.
For the first particle
associated with the undeformed seed operator $V_{(1)}^\mu (k_1)$,
we simply apply the transformation matrix $C^A{}_{\tilde{a}}$ to rotate
$q$-orthonormal directions to $k_1$ transverse directions.

\paragraph{Exact identity in terms of standard scattering amplitudes}
\label{sec:exact-identity-terms-2}

We can now write down Moore's identity \eqref{eq:br_rel_2000-1100}
in terms of the amplitudes in the scattering helicity basis as
\begin{align}
&
\sum_{a',b'}  G^{AB}_{a'b'} \mathcal{T}_{[2000]}^{a'b'}
+G^{AB} \sum_{a'} \mathcal{T}_{[2000]}^{a'a'}
=
\sum_{\tilde a, \tilde b}  C^A{}_{\tilde a} \tilde{G}^{B}_{\tilde b} \mathcal{T}_{[1100]}^{\tilde a|\tilde b}
\,.
\end{align}
where 
$a',b'=L, T,I,J_i$ and $\tilde{a},\tilde{b}=T_R,I_R,J_i$.
On the left hand side, the amplitudes with a completely transverse
polarization ($e^I$ or $e^{J_i}$) are trivially zero,
while on the right hand side, completely transverse directions
should appear in a pairwise way for the amplitude to be non-vanishing.
Therefore, the summation can be made explicit as
\begin{align}
&
\big(  G^{AB}_{TT}+G^{AB} \big) \mathcal{T}_{[2000]}^{TT}
+2G^{AB}_{LT} \mathcal{T}_{[2000]}^{LT}
+\big( G^{AB}_{LL}+G^{AB} \big) \mathcal{T}_{[2000]}^{LL}
\nn\\=&
  C^A{}_{T_R} \tilde{G}^{B}_{T_R} \mathcal{T}_{[1100]}^{T_R|T_R}
+  C^A{}_{I_R} \tilde{G}^{B}_{I_R} \mathcal{T}_{[1100]}^{I_R|I_R}
+\sum_i  C^A{}_{J_i} \tilde{G}^{B}_{J_i} \mathcal{T}_{[1100]}^{J_i|J_i}
\label{eq:2000-1100_cov_rel2}
\,.
\end{align}
In this expression, the coefficients $G$, $C$, and $\tilde{G}$ have
nontrivial $s$ and $t$ dependence to make the equality hold.
As we shall see in Section~\ref{sec:HE_expansion}, the high-energy expansion of
these
coefficients gives proportional constants of linear relations in
the fixed-angle high-energy limit.

\paragraph{The exact identity in terms of DDF amplitudes}
\label{sec:ddf-stat-ampl}

As explained in Appendix \ref{sec:from-covar-posit},
the bracket state \eqref{eq:br_2000}
can further be transformed into the sum of the DDF states,
up to zero norm states, as
\begin{align}
  \eqref{eq:br_2000} =&
\sum_{a,b} D^{AB}_{ab} \ket{ab;\tilde{k}_1}_{DDF}
+\sum_a D^{AB}_a \ket{a;\tilde{k}_1}_{DDF}
+D^{AB} \sum_a \ket{aa;\tilde{k}_1}_{DDF}
% \nn\\&
% +(\text{zero norm states}) 
\,,\nn
%\label{eq:gen_lev2_DDF}
\end{align}
where
$a,b=T,I,J_i$, the transverse directions,
and 
$|\cdots ;\tilde{k}_1 \rangle_{DDF}$ are the DDF states 
given by the action of DDF raising operators $A_{-n}^a$
on a tachyonic vacuum.
%See Appendix \ref{sec:ddf-states} for complete description.
The coefficients are given as
\begin{align}
D^{AB}_{ab}=& G^{AB}_{ab}
\,,
\qquad
D^{AB}_a=
-\sqrt{2}G^{AB}_{L a}
\,,\qquad
D^{AB}=
\frac{1}{4}\big( G_{LL}^{AB}  + 5G^{AB} \big)
\,.
\end{align}
For the massless bracket state 
Since there is no distinction between positive norm
states and DDF states for massless states,
in $\mathcal{A}[1\tilde{1}00]$ side
we use \eqref{eq:18}
to write down Moore's relation in terms of DDF amplitudes,
\begin{align}
&
\sum_{a,b}  D^{AB}_{ab} \mathcal{T}_{DDF[2000]}^{ab}
+\sum_a D^{AB}_{a} \mathcal{T}_{DDF[2000]}^{a}
+D^{AB} \sum_{b} \mathcal{T}_{DDF[2000]}^{bb}
\nn\\=&
\sum_{\tilde a, \tilde b}  C^A{}_{\tilde a} \tilde{G}^{B}_{\tilde b}
\mathcal{T}_{[1100]}^{\tilde a|\tilde b}
\,,
\end{align}
where $a,b=T,I,J_i$
and $\mathcal{T}_{DDF[2000]}^{ab}$ is defined as an amplitude
with one DDF state $|ab; \tilde{k}_1\rangle_{DDF}$ and three tachyons.
The others are understood in the same manner.
One can again use the non-vanishing conditions for the choice of 
completely transverse directions, to make summation explicit as
\begin{align}
&
  D^{AB}_{TT} \mathcal{T}_{DDF[2000]}^{TT}
+  D^{AB}_{II} \mathcal{T}_{DDF[2000]}^{II}
+\sum_i  D^{AB}_{J_iJ_i} \mathcal{T}_{DDF[2000]}^{J_iJ_i}
+D^{AB}_{T} \mathcal{T}_{DDF[2000]}^{T}
\nn\\&\hskip2em
+D^{AB}
\big( \mathcal{T}_{DDF[2000]}^{TT}
+ \mathcal{T}_{DDF[2000]}^{II}+22 \mathcal{T}_{DDF[2000]}^{JJ} \big)
\nn\\=&
  C^A{}_{T_R} \tilde{G}^{B}_{T_R} \mathcal{T}_{[1100]}^{T_R|T_R}
+  C^A{}_{I_R} \tilde{G}^{B}_{I_R} \mathcal{T}_{[1100]}^{I_R|I_R}
+\sum_i  C^A{}_{J_i} \tilde{G}^{B}_{J_i} \mathcal{T}_{[1100]}^{J_i|J_i}
\,,
\label{eq:DDF_rel2_2000-1100}
\end{align}
where, for the third term in the second line,
all $J_i$ ($i=1,\cdots,22$) gives the same result, and then we can take
$J$ as a representative of $J_i$ and multiply 22.

\subsection{Exact identities for Case study I{}I}
\label{sec:bracket-states-terms-3000-1200}

Here we discuss the bracket states that appear in the discussion  of ${\cal A}[\tilde{3}000]={\cal A}[1\tilde{2}00]$ relation,
 presented in Section \ref{sec:2}.
As in the previous case, we will
decompose the relevant bracket states in terms of standard
positive norm states as well as DDF states.
The discussion is parallel to the previous subsection,
and the readers who do not need the details can skip to
Section \ref{sec:HE_expansion}.

We again need to prepare polarization tensors of the seed and the deformer
operators to make the corresponding bracket states  physical.
We set both seed and deformer
operators to be physical.
For the seed operator, we impose
\begin{align}
 \zeta_{1\mu} = e^C_\mu \,,
\qquad
 e^C \in e^{T_q} ,e^{I_q},  e^{J_i} \,.
\end{align}
The polarization tensors of the deformer operator
that satisfy the physical state conditions are
\begin{align}
  \begin{cases}
\text{(I)}:\quad  \zeta_{q\mu\nu} = e^{AB}_{(\mu\nu)}
-\frac{\delta^{AB}}{24}E_{\mu\nu}
 \,,
  \\
\text{(II)}:\quad  \zeta_{q\mu\nu} %= e^{PA}_{(\mu\nu)}+3e^{QA}_{(\mu\nu)}
= 2\sqrt{2} e^{L_q A}_{(\mu\nu)}
 \,,
\\
\text{(III)}:\quad
\zeta_{q\mu\nu} =
24 e^{L_q L_q}_{\mu\nu}
-E_{\mu\nu}
 \,,
  \end{cases}
\qquad
\bigg( E_{\mu\nu} \equiv \sum_{D=T_q, I_q, J_i} e^D_\mu e^D_\nu \bigg)
\end{align}
where
$A,B=T_q, I_q, J_i$
and $\tilde\zeta_{q\mu}=0$ for all the cases.
 $e^{L_q}=(e^P + 3 e^Q)/2\sqrt{2}$ is the longitudinal
polarization with respect to $q$; namely, $e^{P_q}=q/\sqrt{2}$, $e^{L_q}$,
$e^{T_q}$, $e^{I_q}$, and $e^{J_i}$ form a  helicity basis
with respect to $q$.
In this case, the bracket state polarization tensors depend on different
numbers of $q$-transverse directions, and
we will call these three cases Choice (I), (II), and (III)
respectively.
By plugging these in \eqref{eq:3000-zeta_1}--\eqref{eq:1200_zetaR_2},
one obtains the polarization tensors for the bracket operators
(the explicit forms are given later), and by using them
the exact bracket relations for a given set of $A,B,C$
are simply written as
\begin{align}
  \label{eq:78}
  \mathcal{T}_{\text{br}[\tilde{3}000]}^{ABC} =& \mathcal{T}_{\text{br}[1\tilde{2}00]}^{C|AB} 
\,,\qquad
  \mathcal{T}_{\text{br}[\tilde{3}000]}^{AC} = \mathcal{T}_{\text{br}[1\tilde{2}00]}^{C|A} 
\,,\qquad
  \mathcal{T}_{\text{br}[\tilde{3}000]}^{C} = \mathcal{T}_{\text{br}[1\tilde{2}00]}^{C} 
\,,
\end{align}
for Choice (I), (II), and (III), respectively.
As in the previous example, these relations can be
represented as a relation among scattering amplitudes in
$q$-orthonormal basis, and one can check that again the coefficients
are very simple $t$-independent ones.
We do not spell out them here and directly move on to scattering
helicity basis expressions.

\paragraph{Level 3 bracket state in $\mathcal{A}[\tilde{3}000]$ amplitude}
\label{sec:level-3-bracket}

We start with Choice (I).
With this choice, the polarization tensors for the bracket states are
\begin{align}
  \zeta_{\mu\nu\rho}^{(3)} =&
e^{ABC}_{(\mu\nu\rho)}
+\frac{1}{2} q_{(\mu} q_\nu (\delta^{AC} e^B + \delta^{BC} e^A)_{\rho)}
-\frac{\delta^{AB}}{24} \left[
E_{(\mu\nu} e^C_{\rho)}
+q_{(\mu} q_\nu e^C_{\rho)}
\right]
\,,\\
\zeta_{\mu; \nu}^{(3)} =&
(\delta^{AC} e^B + \delta^{BC} e^A)_{\mu}q_\nu
+\frac{1}{2} q_\mu (\delta^{AC} e^B + \delta^{BC} e^A)_{\nu}
-\frac{\delta^{AB}}{24} \left[
2e^C_{\mu}q_{\nu} + q_{\mu} e^C_{\nu}
\right]
 \,,\\
\zeta^{(3)}_\mu =& 
(\delta^{AC} e^B + \delta^{BC} e^A)_{\mu}
-\frac{2\delta^{AB}}{24} e^C_\mu
 \,.
\end{align}
The state corresponding to 
$V_{(3)}^\text{br}(\tilde{k}_1,z;\zeta^{(3)})$
is rewritten in terms of scattering helicity basis
as before,
\begin{align}
  &  \ket{\zeta_{q}=e^{AB}-\frac{\delta^{AB}}{24}E,\tilde\zeta_q=0, \zeta_1=e^C
;\tilde{k}_1}_\text{br}
\nn\\=&
\bigg[
\alpha_{-1}^{ABC}
%\nn\\& \hskip1em
+\delta^{AC}
 \bigg( \frac{1}{2}\alpha_{-1}^{qqB} 
+ \alpha_{-2}^B \alpha_{-1}^q + \frac{1}{2} \alpha_{-2}^q \alpha_{-1}^B
+\alpha_{-3}^B \bigg) +(A \leftrightarrow B)
\nn\\&\hskip1em
-\frac{\delta^{AB}}{24} \bigg(
\sum_D  \alpha_{-1}^{DDC}
+\alpha_{-1}^{qqC}
+2 \alpha_{-2}^C \alpha_{-1}^q
+\alpha_{-2}^q \alpha_{-1}^C
+2 \alpha_{-3}^C
\bigg)
 \bigg] \ket{0;\tilde{k}_1}
\nn\\=&
\bigg[
\sum_{a',b',c'} G^{ABC}_{a'b'c'} \alpha_{-1}^{a'b'c'}
+\sum_{a',b'} G^{ABC}_{[a'b']}\alpha_{-2}^{[a'} \alpha_{-1}^{b']}
 \bigg] \ket{0;\tilde{k}_1}
%\nn\\&
 +(\text{zero norm states})
\,,
\label{bracket_decomp_lv3_1}
\end{align}
where $a', b'=L,T,I,J_i$, namely the longitudinal and
the transverse directions with respect to
$\tilde{k}_1$.
The coefficients are
\begin{align}
    G^{ABC}_{a'b'c'}=&
\bigg[ C^{A} C^B C^{C}
+\frac{\delta^{AC}}{24}
\big(2 C^Q C^Q C^B
-\sum_D C^D C^D C^B \big)
 +(A \leftrightarrow B)
\nn\\&
-\frac{\delta^{AB}}{288}
\big(2  C^Q C^Q C^C
+11\sum_D C^D C^D C^C \big)
\bigg]_{(a'b'c')}
\,,
\\
G^{ABC}_{[a'b']}=&
\bigg[ -\frac{\delta^{AC}}{4} C^{Q} C^{B}
+(A \leftrightarrow B)
+\frac{\delta^{AB}}{48} C^{Q} C^{C}
\bigg]_{[a'b']}
\label{eq:cov_Cho_I}
\,.
\end{align}
$C^A{}_{a}=e^A \cdot e^a$
 is the transformation matrix from $q$-basis
to $\tilde{k}_1$-basis as before.

For Choice (II) and (III), we can repeat the same procedure
and just display the results here.
For Choice (II),
the polarization tensors are
\begin{align}
  \zeta^{(3)}_{\mu\nu\rho}=& 
-2 e^{(P-Q)AC}_{(\mu\nu\rho)}
+\frac{\delta^{AC}}{3} \left(
9e^{PPQ} +6e^{PQQ} + e^{QQQ} \right)_{(\mu\nu\rho)}
\,,\\
\zeta^{(3)}_{\mu; \nu} =&
-2e^A_{(\mu} e^C_{\nu)}
-2e^A_{[\mu} e^C_{\nu]}
+2\delta^{AC} \left(
3e^P_{(\mu} e^Q_{\nu)}
-e^P_{[\mu} e^Q_{\nu]}
+e^Q_\mu e^Q_\nu
\right)
\,,\\
\zeta^{(3)}_\mu =&
 \frac{8\delta^{AC}}{3} e^Q_\mu \,,
\end{align}
and the bracket state is decomposed in the same way as
\eqref{bracket_decomp_lv3_1} with $G^{ABC}_{a'b'c'}$ and
$G^{ABC}_{[a'b']}$ replaced with
\begin{align}
G^{AC}_{a'b'c'}=&
\bigg( 2 C^A C^C C^Q + \frac{2\delta^{AC}}{9}C^Q C^Q C^Q
-\frac{\delta^{AC}}{9}\sum_D  C^D C^D C^Q 
\bigg)_{(a'b'c')} \,,
\\
G^{AC}_{[a'b']} =& -2 C^A{}_{[a'} C^C{}_{b']}
\,.
\label{G_AC}
\end{align}
For Choice (III),
the polarization tensors are
\begin{align}
    \zeta^{(3)}_{\mu\nu\rho}=& 
\frac{1}{4} \left( \left[
-15 e^{PP} -138 e^{PQ} + 41 e^{QQ} \right] e^C \right)_{(\mu\nu\rho)}
-\sum_D e^{DDC}_{(\mu\nu\rho)} \,,
\\
\zeta^{(3)}_{\mu ; \nu} =& 
-\frac{1}{2} e^{(9P+67Q)}_\mu e^C_\nu 
-e^C_\mu e^{(3P+Q)}_\nu 
\,,\\
\zeta^{(3)}_\mu =& -2 e^C_\mu \,,
\end{align}
and the decomposition is carried out with
\begin{align}
    G^{C}_{a'b'c'} =& \frac{5}{12} \bigg(
26 C^Q C^Q C^C - \sum_D C^D C^D C^C \bigg)_{(a'b'c')} \,,\quad
G^C_{[a'b']} = -\frac{65}{2} C^Q{}_{[a'} C^C_{b']} \,.
\label{G_C}
\end{align}

\paragraph{Level 2 bracket state in $\mathcal{A}[1\tilde{2}00]$ amplitude}
\label{sec:level-2-bracket-1}

We move on to the bracket state in ${\cal A}[1\tilde{2}00]$ side.
We first investigate the  Choice (I) case.
The polarization tensors are
\begin{align}
\zeta_{R \mu\nu} =&
e^{AB}_{(\mu\nu)}
+q_{(\mu} (k_2^A e^B + k_2^B e^A)_{\nu)}
+\frac{k_2^A k_2^B }{2} q_\mu q_\nu
\nn\\&
-\frac{\delta^{AB}}{24} \left[
E_{\mu\nu} + 2\sum_D q_{(\mu} e^D_{\nu)} k_2^D
+\frac{\sum_D (k_2^D)^2}{2}q_\mu q_\nu
\right]
\,,\\
\zeta_{R\mu} =&   (k_2^A e^B + k_2^B e^A)_\mu
+\frac{k_2^A k_2^B }{2} q_\mu 
-\frac{\delta^{AB}}{24} \left[
2\sum_D e^D_{\mu} k_2^D
+\frac{\sum_D (k_2^D)^2}{2}q_\mu
\right]
\,.
\end{align}

The state corresponding to 
$V_{(2)}^\text{br}(\tilde{k}_2)$
 is
\begin{align}
  &   \ket{\zeta_{q}=e^{AB}-\frac{\delta^{AB}}{24}E ,\tilde\zeta_q=0; \tilde{k}_2}_\text{br}
\nn\\=&
\bigg[\alpha_{-1}^{AB} 
+\bigg( k_2^A \alpha_{-1}^{Bq} %+(A \leftrightarrow B)
+k_2^A \alpha_{-2}^{B} +(A \leftrightarrow B) \bigg)
+\frac{k_2^A k_2^B}{2}
\big( \alpha_{-1}^{qq} + \alpha_{-2}^{q} \big)
\nn\\&
-\frac{\delta^{AB}}{24}\bigg(
\sum_D \alpha_{-1}^{DD}
+2 \sum_D k_2^D \big( \alpha_{-1}^{Dq} +  \alpha_{-2}^D \big)
+\frac{\sum_D (k_2^D)^2}{2} \big( \alpha_{-1}^{qq} + \alpha_{-2}^q  \big)
\bigg)
\bigg] \ket{0;\tilde{k}_2}
\label{eq:rhs_br_Cho_I}
\,.
\end{align}
We decompose this state by use of the helicity states
with respect to
 ${k}_1$,
introduced
in Section \ref{sec:kinem-case-stud}\footnote{
In the $k_1$ helicity basis,
there exist another null vector, may be called
$e^{\tilde L_1}$, which satisfies
$e^{P_1} \cdot e^{\tilde{L}_1}=1$ and is transverse to the other basis vectors.
In the discussion of the physical amplitudes, $e^{\tilde L_1}$ turns out to be
irrelevant to our analysis.},
as
\begin{align}
  \eqref{eq:rhs_br_Cho_I} =&
\bigg[
\sum_{\tilde a', \tilde b'}
\tilde G^{AB}_{\tilde{a}' \tilde{b}'} \alpha_{-1}^{\tilde{a}' \tilde{b}'}
+\tilde G \sum_{\tilde{a}'} \alpha_{-1}^{\tilde{a}' \tilde{a}'}
\bigg] \ket{0;\tilde{k}_2}
%\nn\\&
+(\text{zero norm states})
\label{eq:rhs_cov_Cho_I}
\,,
\end{align}
where $\tilde{a}',\tilde{b}' =L_2,  T_R, I_R, J_i$, namely the longitudinal
and the transverse directions
with respect to ${k}_1$.
The coefficients are
\begin{align}
  \label{eq:51}
  \tilde G^{AB}_{\tilde{a}' \tilde{b}'} =&
\bigg\{
\bigg( C^A C^B - \frac{\delta^{AB}}{24}\sum_D C^D C^D \bigg)
+2 \bigg( k_2^{(A} C^{B)}- \frac{ \delta^{AB}}{24}\sum_D k_2^D C^D \bigg)
 c
\nn\\& \hskip1em
+\frac{1}{2}
\bigg( k_2^A k_2^B - \frac{\delta^{AB}}{24} \sum_D (k_2^D)^2
\bigg) c c
\bigg\}_{(\tilde{a}' \tilde{b}')}
\,,\\
\tilde G^{AB}=&  \frac{1}{20}\bigg( k_2^A k_2^B - \frac{\delta^{AB}}{24} \sum_D (k_2^D)^2
\bigg) \,.
\end{align}
Here, $C^A{}_{\tilde{a}'}$ is defined by $e^A \cdot e^{\tilde{a}'}$
and we do not list up the explicit components here.
The lower case $c_{\tilde{a}'}$ is the $e^{\tilde{a}'}$ component of
$q$.

For Choice (II),
the polarization tensors are
\begin{align}
    \zeta_{R\mu\nu} =& 2\sqrt{2} \left(
e^{L_q A}_{(\mu\nu)}
+k_2^{L_q} q_{(\mu} e^A_{\nu)}
+k_2^{A} q_{(\mu} e^{L_q}_{\nu)}
+\frac{k_2^{L_q}k_2^A}{2} q_\mu q_\nu
\right) \,,\\
\zeta_{R\mu}=&
2\sqrt{2}
\bigg(k_2^{L_q} e^A_{\mu}
+k_2^{A} e^{L_q}_{\mu}
+\frac{k_2^{L_q}k_2^A}{2} q_\mu \bigg)
\,,
\end{align}
and the corresponding bracket state is decomposed 
as in \eqref{eq:rhs_cov_Cho_I} with
\begin{align}
  \label{G_A}
  \tilde G^A_{\tilde{a}' \tilde{b}'} =&
2\sqrt{2} \bigg(
C^{L_q} C^A + k_2^{L_q} C^A c + k_2^A C^{L_q} c + \frac{k_2^A k_2^{L_q}}{2} c c
\bigg)_{(\tilde{a}' \tilde{b}')} \,,
\\
\tilde G^A =& \frac{2\sqrt{2}k_2^A k_2^{L_q}}{20} \,.
\end{align}
For Choice (III),
the polarization tensors are
\begin{align}
  \zeta_{R\mu\nu} =&
24 e^{L_q L_q}_{(\mu\nu)}
-E_{(\mu\nu)}
+48 k_2^{L_q} q_{(\mu} e^{L_q}_{\nu)}
-2 \sum_D k_2^D q_{(\mu} e^D_{\nu)}
+\frac{24(k_2^{L_q})^2-\sum_D (k_2^D)^2}{2} q_\mu q_\nu \,,
\\
\zeta_{R\mu}=&
48 k_2^{L_q} e^{L_q}_{\mu} -2 \sum_D k_2^D e^D_{\mu}
+ \frac{24(k_2^{L_q})^2-\sum_D (k_2^D)^2}{2} q_\mu  
\,,
\end{align}
and the coefficients for the decomposition of the bracket state are
\begin{align}
    \tilde G_{\tilde{a}' \tilde{b}'}=&
\bigg(
24 C^{L_q}{}C^{L_q}{}
-\sum_D C^D{}C^D
+48 k_2^{L_q} C^{L_q} \, c
-2\sum_D k_2^D C^{D} \, c
\nn\\& \hskip1em
+\frac{24(k_2^{L_q})^2-\sum_D (k_2^D)^2}{2}  \, cc
\bigg)_{(\tilde{a}' \tilde{b}')}
\,,\\
\tilde G=&
\frac{24(k_2^{L_q})^2-\sum_D (k_2^D)^2}{20} 
\label{G_none}
\,.
\end{align}
 
\paragraph{The exact identity in terms of the standard scattering amplitudes}
\label{sec:exact-identity-terms-1}

Now we can write down the exact identity relation in terms of
$\mathcal{T}_{[3000]}^{\mu\nu\rho}$, $\mathcal{T}_{[3000]}^{\mu ; \nu}$,
and $\mathcal{T}_{[3000]}^{\mu}$ amplitudes which are ``polynomial
pieces'' of the scattering amplitudes with
$V_{(3)}(\tilde{k}_1,x;\zeta)$
and three tachyons insertion.
For right hand side, the amplitude pieces are denoted as
$\mathcal{T}_{[1200]}^{\mu|\nu\rho}$
and $\mathcal{T}_{[1200]}^{\mu|\nu}$, which come from
a $V_{(1)}(k_1,x;\zeta)$, $V_{(2)}(\tilde{k}_2,0;\zeta)$
and two tachyons amplitude.
Like the previous example, the vertical line in the superscript
separates an index from the first particle from ones of the second.
With these pieces of the amplitudes, the exact identity is given 
as
\begin{align}
  \label{eq:79}
&
\sum_{a',b',c'}  G^{ABC}_{a'b'c'} \mathcal{T}_{[3000]}^{a'b'c'}
+\sum_{a',b'} G^{ABC}_{[a'b']} \mathcal{T}_{[3000]}^{[a';b']}
%\nn\\=&
=
\sum_{\tilde c} C^C{}_{\tilde c}
\bigg(
\sum_{\tilde a', \tilde b'} 
\tilde{G}^{AB}_{\tilde a' \tilde b'} \mathcal{T}_{[1200]}^{\tilde c|\tilde a' \tilde b'}
+  \tilde{G}^{AB} \sum_{\tilde{a}'}
   \mathcal{T}_{[1200]}^{\tilde c|\tilde{a}' \tilde{a}'} \bigg)
\,,\nn
\end{align}
for Choice (I), 
where
the indices are 
$a',b',c'=L,T,I,J_i$,  $\tilde{a}',\tilde{b}'=L_2,T_R,I_R,J_i$, and
$\tilde c=T_R, I_R, J_i$.
Here, $[a;b]$ represents the anti-symmetrization of the indices
and the symmetrized ones are missing since they appear only 
as a part of decoupling amplitudes.
There are also similar relations for Choice (II) and (III),
where $G$ coefficients are replaced with the ones in 
\eqref{G_AC}, \eqref{G_A} and \eqref{G_A}, \eqref{G_none}
respectively.
By dropping trivially vanishing amplitudes summation is made
explicit as,
for Choice (I),
($B$ and $AB$ indices will be missing for Choice (II) and (III)
respectively)
\begin{align}
&
  G^{ABC}_{TTT} \mathcal{T}_{[3000]}^{TTT}
+3  G^{ABC}_{LTT} \mathcal{T}_{[3000]}^{LTT}
+3  G^{ABC}_{LLT} \mathcal{T}_{[3000]}^{LLT}
+  G^{ABC}_{TTT} \mathcal{T}_{[3000]}^{TTT}
+2 G^{ABC}_{[TL]} \mathcal{T}_{[3000]}^{[T;L]}
\nn\\=&
  C^C{}_{T_R}
\bigg(
\big(  \tilde{G}^{AB}_{T_RT_R}  +\tilde{G}^{AB} \big)
\mathcal{T}_{[1200]}^{T_R|T_RT_R}
+2 \tilde{G}^{AB}_{L_2T_R} \mathcal{T}_{[1200]}^{T_R|L_2T_R}
+\big( \tilde{G}^{AB}_{L_2L_2}+\tilde{G}^{AB} \big) \mathcal{T}_{[1200]}^{T_R|L_2L_2}
\bigg)
\nn\\&
+  C^C{}_{I_R} \bigg(
\tilde{G}^{AB}_{T_RI_R} \mathcal{T}_{[1200]}^{I_R|T_RI_R}
+\tilde{G}^{AB}_{L_2I_R} \mathcal{T}_{[1200]}^{I_R|L_2I_R}
\bigg)
\nn\\&
+\sum_{i=1}^{22}  C^C{}_{J_i} \bigg(
\tilde{G}^{AB}_{T_RJ_i} \mathcal{T}_{[1200]}^{J_i|T_RJ_i}
+\tilde{G}^{AB}_{L_2J_i} \mathcal{T}_{[1200]}^{J_i|L_2J_i}
\bigg)
\,.
\end{align}

\paragraph{The exact identity in terms of DDF amplitudes}
\label{sec:exact-identity-terms}

As before, we rewrite the level 3 and 2 bracket
states for Choice (I) in terms
of DDF amplitudes as, up to zero norm states,
\begin{align}
\eqref{eq:cov_Cho_I}
=&  \sum_{a,b,c} D^{ABC}_{abc} \ket{abc}_{DDF}  
+\sum_{a,b} \bigg( D^{ABC}_{(ab)} \ket{(a;b)}_{DDF}
+D^{ABC}_{[ab]} \ket{[a;b]}_{DDF} \bigg)
\nn\\&
+\sum_a  D^{ABC}_{1a} \ket{a}_{DDF}
+\sum_{a,b} D^{ABC}_{2a}  \ket{abb}_{DDF}
+D^{ABC} \sum_b \ket{b;b}_{DDF}
\,, \nn
\\
  \eqref{eq:rhs_cov_Cho_I} =&
\sum_{\tilde a, \tilde b} \tilde D^{AB}_{\tilde{a}\tilde{b}} \ket{\tilde{a} \tilde{b}}_{DDF}
+\sum_{\tilde a}  \tilde D^{AB}_{\tilde{a}} \ket{\tilde{a}}_{DDF}
+ \tilde D^{AB} \sum_{\tilde{a}} \ket{\tilde{a}\tilde{a}}_{DDF}
 \,, \nn
\end{align}
where $a,b,c=T,I,J_i$ and $\tilde{a},\tilde{b} = T_R, I_R, J_i$.
The coefficients are determined by general consideration in
Appendix \ref{sec:from-covar-posit} as
\begin{align}
  \label{eq:44}
  D^{ABC}_{abc}=& G^{ABC}_{abc} 
\,,\quad
D^{ABC}_{(ab)} =-3G^{ABC}_{Lab} \,,
\quad
D^{ABC}_{[ab]} = G^{ABC}_{[ab]} \,,
\quad
D^{ABC} = -\frac{1}{2}G^{ABC}_{LLL} \,,
\nn\\
D^{ABC}_{1a} =& \frac{1}{4} \left( 9 G^{ABC}_{LLa} + 2G^{ABC}_{[La]} \right)
\,,\qquad
 D^{ABC}_{2a} = \frac{1}{8} \left( 3 G^{ABC}_{LLa} - 2G^{ABC}_{[La]} \right)
\,,
\end{align}
and
\begin{align}
  \label{eq:53}
  \tilde D^{AB}_{\tilde{a}\tilde{b}} =& 
\tilde G^{AB}_{(\tilde{a}\tilde{b})}
\,, \qquad
  \tilde D^{AB}_{\tilde{a}} =
-\sqrt{2} \tilde G^{AB}_{L_2 \tilde{a}}
\,,\qquad
\tilde D^{AB}= \frac{1}{4}\big( \tilde G^{AB}_{L_2 L_2} +5 \tilde G^{AB} \big) \,,
\end{align}
For Choice (II) and (III), we simply replace $G$ coefficients
in the $D$ coefficients with
the corresponding ones.

By use of the DDF amplitudes that correspond to these states,
the exact identities are expressed,
after taking the trivially vanishing amplitudes
into account, as
\begin{align}
  \label{eq:5}
&
\big(  D^{ABC}_{TTT} +D^{ABC}_{2 T} \big) \mathcal{T}_{DDF[3000]}^{TTT}
+ \big(3  D^{ABC}_{TII} +D^{ABC}_{2 T} \big) \mathcal{T}_{DDF[3000]}^{TII}
\nn\\&
+\sum_{i=1}^{22} \big(3  D^{ABC}_{TJ_iJ_i} +D^{ABC}_{2 T} \big) \mathcal{T}_{DDF[3000]}^{TJ_iJ_i}
\nn\\&
+\big( D^{ABC}_{(T;T)} + D^{ABC} \big) \mathcal{T}_{DDF[3000]}^{(T;T)}
+\big( D^{ABC}_{(I;I)} + D^{ABC} \big) \mathcal{T}_{DDF[3000]}^{(I;I)}
\nn\\&
+\sum_{i=1}^{22} \big( D^{ABC}_{(J_i;J_i)} + D^{ABC} \big) \mathcal{T}_{DDF[3000]}^{(J_i;J_i)}
+D^{ABC}_{T} \mathcal{T}_{DDF[3000]}^{T}
\nn\\=&
  C^C{}_{T_R}
\bigg(
 \big( \tilde{D}^{AB}_{T_RT_R}  + D^{AB} \big)
              \mathcal{T}_{DDF[1200]}^{T_R|T_RT_R}
+ \big( \tilde{D}^{AB}_{I_RI_R}  + D^{AB} \big)
              \mathcal{T}_{DDF[1200]}^{T_R|I_RI_R}
\nn\\&\hskip3em
+\sum_{i=1}^{22} \big( \tilde{D}^{AB}_{J_iJ_i}  + D^{AB} \big)
              \mathcal{T}_{DDF[1200]}^{T_R|J_iJ_i}
+\tilde{D}^{AB}_{T_R} \mathcal{T}_{[1200]}^{T_R|T_R}
 \bigg)
\nn\\&
+ C^C{}_{I_R} \bigg( 2\tilde{D}^{AB}_{T_RI_R} \mathcal{T}_{DDF[1200]}^{I_R|T_RI_R}
+\tilde{D}^{AB}_{I_R} \mathcal{T}_{[1200]}^{I_R|I_R} \bigg)
\nn\\&
+\sum_{i=1}^{22} C^C{}_{J_i} \bigg( 2\tilde{D}^{AB}_{T_RJ_i} \mathcal{T}_{DDF[1200]}^{J_i|T_RJ_i}
+\tilde{D}^{AB}_{J_i} \mathcal{T}_{[1200]}^{J_i|J_i} \bigg)
\,.  
\end{align}

\section{High-energy stringy symmetry v.s. exact identities from bracket algebra}
\label{sec:HE_expansion}

In this section, we consider the high-energy expansion of bracket
relations and examine how these relations constrain the asymptotic forms of
scattering amplitudes.

The relations are, for example in the case of
$\mathcal{A}[\tilde{2}000]=\mathcal{A}[1\tilde{1}00]$,
\begin{align}
   \sum_{a',b'} G^{AB}_{a'b'} \mathcal{T}_{[2000]}^{a'b'}
+G^{AB} \sum_{a'} \mathcal{T}_{[2000]}^{a'a'}
=&
\sum_{\tilde a, \tilde b}
  C^A{}_{\tilde a} \tilde{G}^{B}_{\tilde b} \mathcal{T}_{[1100]}^{\tilde a|\tilde b}
\,,\nn
\end{align}
for standard scattering amplitudes.
We expand the transformation matrices, $G$ and $C$, as well as
the amplitudes under the $s\rightarrow \infty$ with $\hat{t}=t/s$ fixed limit.
At each order of $s$, there will be relations among asymptotic amplitudes.
We will explore how these relations ``bootstrap'' asymptotic amplitudes,
 and
whether or not they reproduce known high-energy relations.

In this program, the transformation matrices $C^A{}_a$ (therefore, $G$
and $D$) are regarded as inputs, since they are determined once we
specify the momenta $k_i$ and $q$.
On the other hand, the amplitudes are considered to be unknowns
which are to be determined.
However, we need to supply information on the leading power of each
amplitude\footnote{%
As mentioned in Sections \ref{sec:2} and \ref{sec:bracket-states-terms-2000-1100},
we consider amplitudes up to a common exponential part
($F_{s-t}$ in \eqref{F_s-t}),
and mean the leading power by the leading power of
the rest of ``polynomial'' parts.},
such as $\mathcal{T}_{[2000]}^{TT}=\mathcal{O}(s^3)$.
For both types of amplitudes, a ``power counting rule''
 has been established \cite{Chan:2005ji,Ho:2006zu,DDF-HE} and
 it tells the relative power of a given amplitude
with respect to the leading order power.
Though it turns out that it is actually sufficient to know the relative powers
to carry out the program, to make expressions concrete
we employ our empirical knowledge on the orders of scattering amplitudes.
We also need to use triviality of amplitudes, like completely
transverse directions $J_i$ must appear in a pairwise way for an
amplitude to be non-vanishing.
This fact reduces the number of independent unknowns.
We have already taken this fact into account, for example, in
\eqref{eq:2000-1100_cov_rel2} and \eqref{eq:DDF_rel2_2000-1100}.

\subsection{High-energy linear relations from the decoupling of
high-energy  zero-norm states and saddle-point calculation}
\label{sec:high-energy-linear}

We are about to investigate how
Moore's relations restrict amplitudes under
the fixed-angle high-energy limit.
In order to have a view on what kinds of relations we expect to see, we
briefly review an approach based on the decoupling of (high-energy)
zero-norm states and collect some known linear relations
from \cite{Chan:2003ee,Chan:2004yz,Chan:2005ji}.
We will also mention a couple of relations which are obtained 
by saddle-point calculation.
In the following subsections, we examine which of these relations are
extracted from exact relations by a high-energy expansion.

To illustrate the analysis,
we take four point amplitudes with one level 2 state
and three tachyons, $\mathcal{T}^{\mu\nu}_{[2000]}$, as an example.
For simplicity, the helicity basis with respect to the momentum
for the level 2 state is denoted as $e^P$, $e^L$, and $e^T$ in this
subsection.
The completely transverse directions are irrelevant here.
From the oscillator expressions of zero norm states,
\eqref{eq:ZN_lv2_2} and \eqref{eq:ZN_lv2_1L},
one can immediately see that
the amplitudes obey
the following relations,
\begin{align}
  \label{eq:49}
 5 \mathcal{T}_{[2000]}^{PP}
+ \mathcal{T}_{[2000]}^{LL}
+ \mathcal{T}_{[2000]}^{TT}
+5\sqrt{2} \mathcal{T}_{[2000]}^{P}
=0 \,,\quad
\sqrt{2} \mathcal{T}_{[2000]}^{PL}
+ \mathcal{T}_{[2000]}^{P} =0 \,.
\end{align}
In the high-energy limit, the masses are negligible and
$e^P$ approximates to $e^L$ (as explicitly seen from
\eqref{eq:eP_2000} and \eqref{eq:eL_2000}).
By taking a linear combination, one finds
that in a linear combination
$  \mathcal{T}_{[2000]}^{TT} - 4 \mathcal{T}_{[2000]}^{LL}$
the leading order part vanishes, since this combination approximates
a zero-norm state in the high-energy limit.
One thus obtains a linear relation in the high-energy limit,
\begin{align}
  \label{eq:52}
  \mathcal{T}_{[2000]}^{TT} = 4 \mathcal{T}_{[2000]}^{LL}
\,.
\end{align}
Actually, in order to come to this conclusion, one need to be sure
that these two are indeed of leading order.
As mentioned in the previous subsection, a ``power counting rule''
of \cite{Chan:2003ee,Chan:2005ji}
tells the relative power of an amplitude with a set of helicity projections
compared to the leading order power,
and the amplitudes in \eqref{eq:52} are
indeed the leading order ones.
Thus this is a high-energy linear relation of this amplitude.
The same argument leads to a linear relation for
$\mathcal{T}_{[3000]}^{\mu\nu\rho}$ \cite{Chan:2004yz},
\begin{align}
  \label{eq:54}
  \mathcal{T}_{[3000]}^{TTT} :
  \mathcal{T}_{[3000]}^{LLT} :
  \mathcal{T}_{[3000]}^{(L;T)} :
  \mathcal{T}_{[3000]}^{[L;T]} =
8: 1: -1: -1 \,,
\end{align}
where $  \mathcal{T}_{[3000]}^{(L;T)}$ and $  \mathcal{T}_{[3000]}^{[L;T]}$
are symmetric and anti-symmetric combinations of the indices
in an amplitude
that corresponds to $\alpha_{-2}^L \alpha_{-1}^T$.

Such high-energy linear relations based on
the decoupling of zero-norm states should hold in very general
circumstances.
The same relations should hold for other choices of extra vertex
operators (three tachyons in the current examples) and also for all orders in
perturbation theory.
We thus take these relations as symmetry identities in the high-energy
limit.
On the other hand, this argument is on a state-level analysis and is
confined in a set of states at a fixed level.
The decoupling of high-energy zero-norm states does not give
any inter-level relation, but we expect that there appear several
inter-level relations as well, as all the mass levels are degenerate
in the high-energy limit.
Among many possible inter-level linear relations, we may be interested
in the following relations among all $T$-polarized
amplitudes,
\begin{align}
  \label{eq:29}
\mathcal{T}_{[2000]}^{TT} = \mathcal{T}_{[1100]}^{T|T} 
\,, \qquad
\mathcal{T}_{[3000]}^{TTT}=\mathcal{T}_{[1200]}^{T|TT} \,,
\end{align}
at the leading order.
Here, all the states are generated only by the level 1
oscillator $\alpha_{-1}^T$ with $T$-polarization (with respect to the
momentum of the state on which it acts)
and the total level on the both hands sides are
the same.
As long as these two conditions are met, at the leading order,
the same kind of relations
hold in general; for example,
$\mathcal{T}_{[4000]}^{TTTT}=\mathcal{T}_{[3100]}^{TTT|T} =
\mathcal{T}_{[2110]}^{TT|T|T}=\mathcal{T}_{[1111]}^{T|T|T|T}$ and so on.
These relations can be
derived through direct calculation
by use of the saddle-point approximation \cite{Chan:2003ee}.
We will come back to this partonic behavior of scattering amplitudes
in Section \ref{sec:conclusion}, but in this section, we check if this kind
of relation is also obtained through Moore's relations.

In general, the appearance of such leading order relations implies
that it is possible to choose another basis of physical states
such that there exists a unique state in the basis at the leading
order and all the other states are of subleading.
Such basis has been found and discussed in
\cite{Chan:2005ji,Ho:2006zu} and called DDF gauge,
where
positive norm physical states are spanned 
by DDF operators.
The corresponding amplitudes are DDF amplitudes, such as
$\mathcal{T}_{DDF[3000]}^{TTT}$.
In this gauge, the leading energy dependence of
an amplitude is determined by the number of $T$ indices.
For example, at level 3, $\mathcal{T}_{DDF[3000]}^{TTT}$ generated by
$(A_{-1}^T)^3$ starts with the highest power in $s$.
$\mathcal{T}_{DDF[3000]}^{T;T}$ is at the next-to-leading order, and
$\mathcal{T}_{DDF[3000]}^{T}$ and
$\mathcal{T}_{DDF[3000]}^{TII}$ are further sub-leading.
Therefore, in this gauge, leading order linear relations become
trivial, and we can concentrate on inter-level relations like \eqref{eq:29}
as well as \textit{subleading} relations among DDF amplitudes.
Actually, we can develop a systematic high-energy expansion
\cite{DDF-HE}, and observe several interesting relations connecting
amplitudes of different leading energy dependence.
Among amplitudes generated only by $A_{-1}^T$ and $A_{-1}^I$, it is found
\begin{align}
  \label{eq:4}
  \mathcal{T}_{DDF[n000]}^{(T)^{n-m} (I)^m}
=\mathcal{T}_{DDF[n000]}^{(T)^{n-m-2}(I)^{m+2}} \left(
\frac{-2s}{m+1} + \mathcal{O}(s^0) \right) \,.
\end{align}
Namely, up to subleading corrections,
the following relations obey;
\begin{align}
  \label{DDF_HE_relations1}
  \mathcal{T}_{DDF[2000]}^{TT} =-2 s \,
\mathcal{T}_{DDF[2000]}^{II} \,,
\quad
\mathcal{T}_{DDF[4000]}^{TTTT}=-2s \, \mathcal{T}_{DDF[4000]}^{TTII}
= \frac{4s^2}{3} \mathcal{T}_{DDF[4000]}^{IIII} \,.
\end{align}
Since Moore's relation is exact, we should be able to
obtain such inter-level and inter-energy-level relations from it.
As a preliminary trial, we shall derive a first few nontrivial
relations among DDF amplitudes in the following subsections.

\subsection{High-energy expansions of the  scattering amplitudes: ${\cal A}[\tilde{2}000]={\cal A}[1\tilde{1}00]$}

As explained in the beginning of this section,
we take the large-$s$ expansions of the amplitudes and the
coefficients of the exact relation \eqref{eq:2000-1100_cov_rel2},
with $\hat{t}=t/s$ fixed, as
\begin{align}
  \label{eq:6}
  \mathcal{T}_{[2000]}^{TT} =& \mathcal{T}_{[2000](3)}^{TT} s^3 + \mathcal{T}_{[2000](2)}^{TT} s^2 + 
\cdots \,,\\
  G^{T_qT_q}_{TT} =& G^{T_qT_q}_{TT(0)} + G^{T_qT_q}_{TT(-1)} s^{-1} + \cdots,
\end{align}
and so on.
Here, on the right hand side, factors like $\mathcal{T}_{[2000](n)}^{TT}$ denote
 coefficients of $s^n$ and are in general functions of $\hat{t}$, for
 example, $   \mathcal{T}_{[2000](3)}^{TT} =
 \frac{-\hat{t}(1+\hat{t})}{4}$.\footnote{%
Explicit expressions of amplitudes are found in the preprint version
(\texttt{v2}) of the manuscript.
You may obtain all the other coefficients by use of them.}

In the bracket relation \eqref{eq:2000-1100_cov_rel2},
by collecting the terms at the same order in $s$,
we can find several relations among these expansion coefficients.
For example, for $(A,B)=(T_q,T_q)$ choice, the coefficient of $s^3$ reads
\begin{align}
  \label{eq:8}
0=&  \big(G^{T_qT_q}_{TT(0)} + G^{T_qT_q}_{(0)} \big) \mathcal{T}_{[2000](3)}^{TT}
+\big( G^{T_qT_q}_{LL(0)} + G^{T_qT_q}_{(0)} \big) \mathcal{T}_{[2000](3)}^{LL}
- C^{T_q}{}_{T_R(0)} \tilde{G}^{T_q}_{T_R(0)} \mathcal{T}_{[1100](3)}^{T_R|T_R}
\,.
\end{align}
After evaluating $C$ and $G$ coefficients by use of their asymptotic forms,
it leads to asymptotic relations among the leading order
part of scattering amplitudes.

Before going further, we point out that the coefficient function for the right hand side,
$C^A{}_{\tilde{a}}\tilde{G}^B_{\tilde{b}}$ is constant
and almost diagonal.
The amplitudes $\mathcal{T}_{[1100]}^{\tilde a|\tilde b}$ are non-zero
only for $(\tilde{a}, \tilde{b})=(T_R, T_R)$, $(T_R, \tilde{L}_2)$,
$(I_R,I_R)$, and $(J_i, J_i)$.
Here $e^{\tilde L_2}$ is one of the basis vector for $\tilde{k}_2$
 helicity basis, $k_2 \cdot e^{\tilde L_2}=1$. 
For these $\tilde{a}, \tilde{b}$, non-vanishing coefficients are
\begin{align}
  C^{I_q}{}_{I_R}\tilde{G}^{I_q}_{I_R} =-1 \,,
\qquad
  C^{T_q}{}_{T_R}\tilde{G}^{T_q}_{T_R} =1 \,,
\qquad
  C^{J_i}{}_{J_i}\tilde{G}^{J_i}_{J_i} =1
\quad (\text{no sum for $i$})
 \,,\nn
\end{align}
and the unphysical projection onto $e^{\tilde L_2}$
does not appear.
Therefore, on the right hand side, 
the coefficients can be regarded as constants and
the asymptotic expansion only involves
the expansion of the amplitudes.

To first two orders with $(A,B)=(T_q,T_q), (I_q,I_q), (J,J)$,
and $(T_q, I_q)$, the relations are
\begin{align}
% (A,B)=  (T_q, T_q)&: \, 
% \nn\\ 
\mathcal{O}(s^3) : \qquad
0=&
\frac{19}{20}\mathcal{T}^{TT}_{[2000](3)}
+\frac{1}{5} \mathcal{T}^{LL}_{[2000](3)}
-\mathcal{T}^{T_R|T_R}_{[1100](3)}
\label{eq:2000-1100_TT_3}
\,,\\
0=& \mathcal{T}^{TT}_{[2000](3)} -4 \mathcal{T}^{LL}_{[2000](3)}
\label{eq:2000-1100_II_3}
\,,\\
\mathcal{O}(s^2) : \qquad
0=&
\frac{19}{20}\mathcal{T}^{TT}_{[2000](2)}
-2 \mathcal{T}^{LL}_{[2000](3)}
+\frac{1}{5} \mathcal{T}^{LL}_{[2000](2)}
-\mathcal{T}^{T_R|T_R}_{[1100](2)}
\label{eq:2000-1100_TT_2}
\,,\\
0=&
-\frac{1}{20} \mathcal{T}^{TT}_{[2000](2)}
+6 \mathcal{T}^{LL}_{[2000](3)}
+\frac{1}{5} \mathcal{T}^{LL}_{[2000](2)}
+\mathcal{T}^{I_R|I_R}_{[1100](2)}
\,,
\label{eq:2000-1100_II_2}
\\
0=&
-\frac{1}{20} \mathcal{T}^{TT}_{[2000](2)}
-2 \mathcal{T}^{LL}_{[2000](3)}
+\frac{1}{5} \mathcal{T}^{LL}_{[2000](2)}
-\mathcal{T}^{J|J}_{[1100](2)}
\,,
\label{eq:2000-1100_JJ_2}
\\
0=&
(2\hat{t}+1) \mathcal{T}^{TT}_{[2000](3)}
+\sqrt{-2\hat{t}(1+\hat{t})} \mathcal{T}^{TL}_{[2000](5/2)}
\,,
\label{eq:2000-1100_TL_2}
\\
\mathcal{O}(s) : \qquad 
0=&
-2(4\hat{t}^2+6\hat{t}+1) \mathcal{T}^{TT}_{[2000](3)}
+(2\hat{t}^2 + 3\hat{t} +1) \mathcal{T}^{TT}_{[2000](2)}
\nn\\&
+\sqrt{-2\hat{t}(1+\hat{t})} (1+\hat{t}) \mathcal{T}^{TL}_{[2000](3/2)}
\label{eq:2000-1100_TL_1}
\,.
\end{align}
From \eqref{eq:2000-1100_TT_3} and  \eqref{eq:2000-1100_II_3},
it is easy to obtain the following
linear relations
\begin{align}
  \mathcal{T}^{TT}_{[2000](3)} =4 \mathcal{T}^{LL}_{[2000](3)} \,,
\qquad
\mathcal{T}^{TT}_{[2000](3)} = \mathcal{T}_{[1100](3)}^{T_R|T_R} \,.
\label{eq:2000_DDF_leading}
\end{align}
The first one is the linear relation \eqref{eq:52}
derived from the decoupling
of zero-norm states, while the second one is 
an inter-level relation \eqref{eq:29}.

On ${\cal A}[1\tilde{1}00]$ side, 
$I_R$ and $J_i$ have the same geometrical meaning; namely,
both represent completely transverse directions to the scattering plane.
Therefore, replacing $(J_i, J_i)$ with $(I_R, I_R)$ does not change the amplitudes.
More explicitly, $\mathcal{T}^{I_R|I_R}_{[1100]} = \mathcal{T}^{J|J}_{[1100]}$
holds for all orders in $s$.
By use of this rotational symmetry
for \eqref{eq:2000-1100_TT_2},
\eqref{eq:2000-1100_II_2}, and
\eqref{eq:2000-1100_JJ_2},
we find further relations,
\begin{align}
  4\mathcal{T}^{LL}_{[2000](3)} +\mathcal{T}^{I_R|I_R}_{[1100](2)} =0 \,,
\quad
\mathcal{T}^{TT}_{[2000](2)}
-\mathcal{T}^{T_R|T_R}_{[1100](2)} +  \mathcal{T}^{I_R|I_R}_{[1100](2)} =0 \,.
\end{align}
The rest of the relations \eqref{eq:2000-1100_TL_2}
and \eqref{eq:2000-1100_TL_1} provide some angle-dependent relations.
As already mentioned,
we have used the explicit $s$ dependence of each amplitudes.
The power counting rule \cite{Chan:2005ji} provides the information
on relative power of each amplitudes.
Since the relation is homogeneous,
these relative powers of the amplitudes suffice
to determine high-energy linear relations;
some of them have been derived by use of
the decoupling of high-energy zero norm states
\cite{Chan:2004yz,Chan:2005ji},
and we also find extra subleading relations
and inter-level relations.

\subsection{High-energy expansions of the DDF amplitudes: ${\cal A}[\tilde{2}000]={\cal A}[1\tilde{1}00]$}
\label{sec:HE_1200-1100_DDF}

We move on to exact relations among DDF amplitudes
\eqref{eq:DDF_rel2_2000-1100}.
Again, we expand all the elements that appear in \eqref{eq:DDF_rel2_2000-1100}
with respect to its $s$ dependence,
${\cal T}_{DDF[2000]}^{TT}= {\cal T}_{DDF[2000]\; (3)}^{TT}\, s^3
+\cdots$,
$D^{I_q I_q}_{II} = D^{I_q I_q}_{II \; (0)}+ D^{I_q I_q}_{II \; (-1)}
\, s^{-1}+\cdots$,
and so on.
We then organize the exact relations with respect to its $s$ dependence as
before, with $D$ coefficients evaluated, as\footnote{%
Recall that the right hand side, $\mathcal{A}[1\tilde{1}00]$ side, is 
the same as the previous case.}
\begin{align}
  {\cal O}(s^3):& \qquad
\mathcal{T}_{DDF[2000]\; (3)}^{TT} = {\cal T}_{[1100] (3)}^{T_R|T_R}
\label{eq:2000-1100_DDF_3}
 \,,\\
  {\cal O}(s^{2}):
& \qquad
\mathcal{T}_{DDF[2000] \; (2)}^{TT}
-\frac{1}{2} \mathcal{T}_{DDF[2000] \; (3)}^{TT}
 = {\cal T}_{[1100] \; (2)}^{T_R|T_R }
\label{eq:2000-1100_DDF_TT_2}
 \,,\\
& \qquad
 \mathcal{T}_{DDF[2000] \; (2)}^{JJ}
-\frac{1}{2} \mathcal{T}_{DDF[2000] \; (3)}^{TT}
 = {\cal T}_{[1100] \; (2)}^{J|J } 
\label{eq:2000-1100_DDF_JJ_2}
\,,\\
& \qquad
 \mathcal{T}_{DDF[2000] \; (2)}^{II}
+\frac{3}{2} \mathcal{T}_{DDF[2000] \; (3)}^{TT}
 = - {\cal T}_{[1100] \; (2)}^{I_R|I_R } 
\label{eq:2000-1100_DDF_II_2}
\,,\\
& \qquad
\frac{4\hat{t}+2}{\sqrt{-\hat{t}(\hat{t}+1)}} \mathcal{T}_{DDF[2000] \; (3)}^{TT}
-2 \mathcal{T}_{DDF[2000] \; (5/2)}^{T}
 =0 \,,
\label{eq:2000-1100_DDF_TI_2}
\end{align}
where $J$ represents one of 22 $J_i$,
and the $\mathcal{O}(s^3)$ relation is for $(A,B)=(T_q, T_q)$ while
at $\mathcal{O}(s^2)$ the relations are for
$(A,B)=(T_q,T_q), (J,J), (I_q, I_q)$ and $(T_q, I_q)$ in order.
By using the fact that ${\cal T}_{DDF[2000]}^{JJ}={\cal T}_{DDF[2000]}^{II}$
and ${\cal T}_{[1100]}^{J|J}={\cal T}_{[1100]}^{I_R | I_R}$,
which follows from the rotational symmetry in the directions
transverse to the scattering plane,
one can further derive
\begin{align}
  2{\cal T}_{DDF[2000]\; (2)}^{II} = {\cal T}_{[1100]\; (2)}^{I_R | I_R} \,.
\label{eq:DDF2000_sub1}
\end{align}
The leading order relation \eqref{eq:2000-1100_DDF_3}
is a DDF-amplitude counter part of the relation \eqref{eq:29}.
\eqref{eq:2000-1100_DDF_II_2}, combined with \eqref{eq:DDF2000_sub1},
gives a known DDF amplitude relation \eqref{DDF_HE_relations1}.

At order ${\cal O}(s)$, the relations
involve $\hat{t}$, then the scattering angle.
For example, for $A=B=I_q$, one finds
(we omit the subscript $DDF[2000]$ and $[1100]$ for simplicity)
\begin{align}
  -2\,\hat{t} (\hat{t} +1)
\mathcal{T}^{I_R|I_R}_{(1)}=&
12(2\hat{t}+1) \sqrt {{\hat{t}}+1}\sqrt {-{\hat{t}}}
\mathcal{T}^T_{(5/2)}
-8(2\hat{t}+1)^2
\mathcal{T}^{TT}_{(3)}
\nn\\&
+ \hat{t} (\hat{t}+1)
\left(
-9
\mathcal{T}^{II}_{(2)}
+2\mathcal{T}^{II}_{(1)}
+3\mathcal{T}^{TT}_{(2)}
+66\mathcal{T}^{JJ}_{(2)} \right)
\,.\nn
\end{align}
By taking some linear combinations of four relations at this order, 
we can derive some $\hat{t}$-independent relations, for example,
\begin{align}
    \mathcal{T}^{T_R|T_R}_{(1)} - \mathcal{T}^{I_R|I_R}_{(1)} +2 \mathcal{T}^{J|J}_{(1)}
= \mathcal{T}^{TT}_{(1)}+\mathcal{T}^{II}_{(1)}+2 \mathcal{T}^{JJ}_{(1)} \,.
\end{align}
Again, by employing an obvious symmetry, $I_R \rightarrow J$ and $J \rightarrow I$
on both hands sides,
we have
\begin{align}
    \mathcal{T}^{T_R|T_R}_{(1)} + \mathcal{T}^{I_R|I_R}_{(1)} 
= \mathcal{T}^{TT}_{(1)}+3 \mathcal{T}^{II}_{(1)} \,.  
\end{align}

In summary, by using the power counting rule of the DDF amplitudes,
the bracket relation provides
some of the known asymptotic relations, such as
\eqref{eq:2000-1100_DDF_3}, and also 
several subleading relations.

\subsection{High-energy expansions of the scattering amplitudes: ${\cal A}[\tilde{3}000]={\cal A}[1\tilde{2}00]$}

As in the case of $\mathcal{A}[\tilde{2}000]=\mathcal{A}[1\tilde{1}00]$ relation,
we expand the amplitudes and the $G$ matrices
in the bracket relation \eqref{eq:79}
to derive the high-energy relations.
In this case, there are many relations for each Choice  (I), (II), and
(III) with various choices of $A,B$, and $C$.

In the case of Choice (I), for $(A,B,C)=(T_q, T_q, T_q)$, 
the first two leading order relations read
\begin{align}
  \mathcal{O}\big( s^{13/2} \big): \quad 
0=&
C^{T_q}{}_{T(0)}
\bigg(
\tilde{G}^{T_qT_q}_{T_RT_R(2)}\mathcal{T}_{[1200](9/2)}^{T_R|T_RT_R}+\tilde{G}^{T_qT_q}_{L_2L_2(2)}\mathcal{T}_{[1200](9/2)}^{T_R|L_2L_2}
\bigg)
\,,\nn\\
  \mathcal{O}\big( s^{11/2} \big) : \quad 
0=&
C^{T_q}{}_{T_R(0)}\tilde{G}^{T_qT_q}_{T_R L_2(3/2)}\mathcal{T}_{[1200](4)}^{T_R|T_RL_2}
+C^{T_q}{}_{T_R(0)}\tilde{G}^{T_qT_q}_{T_RT_R(2)}\mathcal{T}_{[1200](7/2)}^{T_R|T_RT_R}
\nn\\&
+C^{T_q}{}_{T(0)}\tilde{G}^{T_qT_q}_{L_2L_2(2)}\mathcal{T}_{[1200](7/2)}^{T_R|L_2L_2}
+C^{T_q}{}_{T_R(-1)}\tilde{G}^{T_qT_q}_{L_2L_2(2)}\mathcal{T}_{[1200](9/2)}^{T_R|L_2L_2}
\nn\\&
+ \left( C^{T_q}{}_{T_R(-1)}G^{T_qT_q}_{(2)}+C^{T_q}{}_{T_R(0)}\tilde{G}^{T_qT_q}_{(1)}
+C^{T_q}{}_{T_R(0)}\tilde{G}^{T_qT_q}_{T_RT_R(1)} \right) \mathcal{T}_{[1200](9/2)}^{T_R|T_RT_R}
\,.\nn
\end{align}
Interestingly, unlike the previous case, some of $\tilde{G}$ coefficients
on the right hand side have positive power in $s$, while the left hand
side coefficients do not.
Thus, first few leading order relations in $s$ only involve amplitudes
from $\mathcal{A}[1\tilde{2}00]$.
We now write down the leading order relations,
with $C$, $G$ and $\tilde{G}$ coefficients evaluated, for various
choices of $(A,B,C)$ as
\begin{align}
0=&
4 \mathcal{T}_{[1200](9/2)}^{T_R|L_2L_2} 
- \mathcal{T}_{[1200](9/2)}^{T_R|T_R T_R} 
\label{Cov3000-1200_13}
\,,\\
0=&
\sqrt {2}\mathcal{T}_{[1200](4)}^{T_R|T_RL_2}-{\frac { \left( 2\,\hat{t}+1 \right) \mathcal{T}_{[1200](9/2)}^{T_R|T_RT_R}}{\sqrt {-\hat{t}}\sqrt {\hat{t}+1}}}
\label{Cov3000-1200_10}
\,,\\
0=&
-3\,\mathcal{T}_{[3000](9/2)}^{[L;T]}+\frac{1}{6}\mathcal{T}_{[3000](9/2)}^{TTT}+{\frac {83}{3}}\,\mathcal{T}_{[3000](9/2)}^{TLL}+2\,{\frac {\sqrt {2} \left( 2\,\hat{t}+1 \right) \mathcal{T}_{[1200](4)}^{T_R|T_RL_2}}{\sqrt {-\hat{t}}\sqrt {\hat{t}+1}}}
\nn\\&
+4\,\mathcal{T}_{[1200](7/2)}^{I_R|T_RI_R}+2\,{\frac { \left( 2\,\hat{t}+1 \right) ^{2}\mathcal{T}_{[1200](9/2)}^{T_R|T_RT_R}}{ \left( \hat{t}+1 \right) \hat{t}}}
\label{Cov3000-1200_9_1}
\,,\\
0=&
18\,\mathcal{T}_{[3000](9/2)}^{[L;T]}-\mathcal{T}_{[3000](9/2)}^{TTT}+26\,\mathcal{T}_{[3000](9/2)}^{TLL}
\label{Cov3000-1200_9_2}
\,,\\
0=&
-2\,\mathcal{T}_{[3000](9/2)}^{[L;T]}+6\,\mathcal{T}_{[3000](9/2)}^{TLL}+\frac{1}{2}{\frac {\sqrt {2} \left( 2\,\hat{t}+1 \right) \mathcal{T}_{[1200](4)}^{T_R|T_RL_2}}{\sqrt {-\hat{t}}\sqrt {\hat{t}+1}}}+\mathcal{T}_{[1200](7/2)}^{I_R|T_RI_R}
\nn\\&
+\frac{1}{2}{\frac { \left( 2\,\hat{t}+1 \right) ^{2}\mathcal{T}_{[1200](9/2)}^{T_R|T_RT_R}}{\hat{t}\, \left( \hat{t}+1 \right) }}
\label{Cov3000-1200_9_3}
\,,\\
0=&
-{\frac { \left( 2\,\hat{t}+1 \right) \mathcal{T}_{[3000](9/2)}^{[L;T]}}{\sqrt {-\hat{t}}\sqrt {\hat{t}+1}}}+\frac{2}{3}\mathcal{T}_{[3000](4)}^{TTL}-{\frac {52}{3}}\,\mathcal{T}_{[3000](4)}^{LLL}+\frac{1}{2}{\frac { \left( 2\,\hat{t}+1 \right) \mathcal{T}_{[3000](9/2)}^{TTT}}{\sqrt {-\hat{t}}\sqrt {\hat{t}+1}}}
\nn\\&
-37\,{\frac { \left( 2\,\hat{t}+1 \right) \mathcal{T}_{[3000](9/2)}^{TLL}}{\sqrt {-\hat{t}}\sqrt {\hat{t}+1}}}+4\,\sqrt {2}\mathcal{T}_{[1200](3)}^{J|JL_2}-4\,{\frac { \left( 2\,\hat{t}+1 \right) \mathcal{T}_{[1200](7/2)}^{J|T_RJ}}{\sqrt {-\hat{t}}\sqrt {\hat{t}+1}}}
\label{Cov3000-1200_8_1}
\,,\\
0=&
\frac{1}{6}\mathcal{T}_{[3000](4)}^{TTL}-\frac{13}{3}\mathcal{T}_{[3000](4)}^{LLL}+\frac{1}{9}{\frac { \left( 2\,\hat{t}+1 \right) \mathcal{T}_{[3000](9/2)}^{TTT}}{\sqrt {-\hat{t}}\sqrt {\hat{t}+1}}}-{\frac {80}{9}}\,{\frac { \left( 2\,\hat{t}+1 \right) \mathcal{T}_{[3000](9/2)}^{TLL}}{\sqrt {-\hat{t}}\sqrt {\hat{t}+1}}}
\nn\\&
+\sqrt {2}\mathcal{T}_{[1200](3)}^{J|JL_2}-{\frac { \left( 2\,\hat{t}+1 \right) \mathcal{T}_{[1200](7/2)}^{J|T_RJ}}{\sqrt {-\hat{t}}\sqrt {\hat{t}+1}}}
\label{Cov3000-1200_8_2}
\,,
\end{align}
where 
$J$ stands for one of $J_i$.

From \eqref{Cov3000-1200_13}--\eqref{Cov3000-1200_10}, we find
\begin{align}
\label{Cov3000-1200_Rel1}
  \mathcal{T}_{[1200](9/2)}^{T_R|T_RT_R}
=&
4\,\mathcal{T}_{[1200](9/2)}^{T_R|L_2L_2}  
=
\frac{\sqrt{2}\sqrt {-\hat{t}}\sqrt {\hat{t}+1}} { 2\,\hat{t}+1}
\mathcal{T}_{[1200](4)}^{T_R|T_RL_2}
\,.
\end{align}
The first equality is the linear relation at level 2
we have seen before, \eqref{eq:52}.
Using this relation,
\eqref{Cov3000-1200_9_1}--\eqref{Cov3000-1200_9_3} lead to
\begin{align}
\label{Cov3000-1200_Rel2}
    \mathcal{T}_{[3000](9/2)}^{TTT}
=8 \mathcal{T}_{[3000](9/2)}^{TLL}
= -8 \mathcal{T}_{[3000](9/2)}^{[L;T]}
= - \mathcal{T}_{[1200](7/2)}^{I_R|T_RI_R} \,.
\end{align}
The first three relations are indeed the linear relation 
\eqref{eq:54}
obtained
by use of the decoupling of zero-norm states.
Since $\alpha_{-2}^{(L} \alpha_{-1}^{T)} \ket{0;\tilde{k}_1}$
appears in zero-norm states,
$\mathcal{T}_{[3000]}^{(L;T)}$ is missing in our computation here.

So far, one can see that the leading order amplitudes,
$\mathcal{T}_{[3000](9/2)}^{TTT}$ and $\mathcal{T}_{[1200](9/2)}^{T_R|T_RT_R}$,
can be regarded as basic ones to represent the rest of amplitudes.
The final two relations,
\eqref{Cov3000-1200_8_1} and \eqref{Cov3000-1200_8_2}, give a single relation
\begin{align}
  0=&
 \mathcal{T}_{[3000](4)}^{TTL}
-26 \mathcal{T}_{[3000](4)}^{LLL}
+6 \sqrt{2} \mathcal{T}_{[1200](3)}^{J|JL_2}
\label{eq:3000-1200_cov_s4_1}
\,,
\end{align}
after using the rotational symmetry, 
$\mathcal{T}_{[1200]}^{J|T_RJ} = \mathcal{T}_{[1200]}^{I_R|T_RI_R}$.
It involves other leading order amplitudes.
By direct computation, one can check that these three coefficients are
proportional to $\hat{t}(\hat{t}+1)(2\hat{t}+1)$, and then they
are actually proportional to one another.

The next-to-leading order relations are very complicated in general.
Among those, the highest order one at
$\mathcal{O}(s^{11/2})$ is found to provide a simple relation
\begin{align}
  0=&
10  \mathcal{T}_{[1200](9/2)}^{T_R|T_RT_R}
-\mathcal{T}_{[1200](7/2)}^{T_R|T_RT_R}
+4 \mathcal{T}_{[1200](7/2)}^{T_R|L_2 L_2}
\,.
\end{align}
Explicit evaluation
shows that 
these three are not proportional to one another, as a function of
$\hat{t}$, unlike the previous example.
Other relations are more involved
and contain other subleading order amplitudes.
We do not present explicit forms of them here.

The leading order relations \eqref{Cov3000-1200_Rel1}
and \eqref{Cov3000-1200_Rel2} reproduce some of the leading order relations
observed in \cite{Chan:2004yz,Chan:2005ji}.
The relation between all $T$-polarized   amplitudes,
$\mathcal{T}_{[3000](9/2)}^{TTT} = \mathcal{T}_{[1200](9/2)}^{T_R|T_RT_R}$,
is not obtained from Moore's relations, up to this order.
If we supply it, 
the leading order relations
up to $\mathcal{O}(s^{9/2})$,
\eqref{Cov3000-1200_Rel1}
and \eqref{Cov3000-1200_Rel2}, are written by one of them.

\subsection{High-energy expansions of the DDF amplitudes: ${\cal A}[\tilde{3}000]={\cal A}[1\tilde{2}00]$}

Finally, we consider asymptotic relations among DDF amplitudes, \eqref{eq:5}.
As before, we  expand the DDF amplitudes and the D matrices,
We list the leading order ones from Choice (I), (II), and (III),
with the same structure ones collected, as 
\begin{align}
0=&
 \frac{5\hat{t}^2+5\hat{t}+1}{(-\hat{t})(1+\hat{t})} {\cal T}^{T_R|T_RT_R}_{DDF[1200](9/2)}
+\frac{2\hat{t}+1}{\sqrt{(-\hat{t})(1+\hat{t})}} {\cal T}^{T_R|T_R}_{DDF[1200](4)}
-2 {\cal T}^{T_R|I_RI_R}_{DDF[1200](7/2)} \,,\\
0=&
 \frac{(2\hat{t}+1)(5\hat{t}^2+5\hat{t}+1)}{(-\hat{t})^{3/2}(1+\hat{t})^{3/2}}
 {\cal T}^{T_R|T_RT_R}_{DDF[1200](9/2)}
+ \frac{(2\hat{t}+1)^2}{(-\hat{t})(1+\hat{t})}{\cal T}^{T_R|T_R}_{DDF[1200](4)}
\nn\\&
-\frac{2(2\hat{t}+1)}{\sqrt{(-\hat{t})(1+\hat{t})}}
\big( 2{\cal T}^{I_R|T_RI_R}_{DDF[1200](7/2)} 
+{\cal
  T}^{T_R|I_RI_R}_{DDF[1200](7/2)}  \big)
-2{\cal T}^{I_R|I_R}_{DDF[1200](3)}
\,,\\
0=&
\frac{2\hat{t}+1}{\sqrt{(-\hat{t})(1+\hat{t})}} {\cal T}^{T_R|T_RT_R}_{DDF[1200](9/2)}
+{\cal T}^{T_R|T_R}_{DDF[1200](4)} 
\,,\\
0=&
-2 {\cal T}^{TTT}_{DDF[3000](9/2)}
+ \frac{(2\hat{t}+1)^2}{(-\hat{t})(1+\hat{t})}
 {\cal T}^{T_R|T_RT_R}_{DDF[1200](9/2)}
+\frac{2\hat{t}+1}{\sqrt{(-\hat{t})(1+\hat{t})}}{\cal T}^{T_R|T_R}_{DDF[1200](4)} 
\nn\\&
-2{\cal T}^{I_R|T_RI_R}_{DDF[1200](7/2)} 
\,,\\
\label{eq:12}
0=&
\big( D^{T_q JJ}_{(0)}+ D^{T_qJJ}_{TTT(0)} \big)
\mathcal{T}^{TTT}_{DDF[3000](9/2)}
 \,,\\
%   {\cal O}(s^{4}):& \quad (I_q, J_i ,J_i), \, (J_i, J_i)
% \nn\\
0=&
-\frac{2\hat{t}+1}{\sqrt{(-\hat{t})(1+\hat{t})}}
\big( \mathcal{T}^{TTT}_{DDF[3000](9/2)} +
\mathcal{T}^{J_i|TJ_i}_{DDF[1200](7/2)} \big)
+2\mathcal{T}^{T;T}_{DDF[3000](4)}
%\nn\\&
-\mathcal{T}^{J_i|J_i}_{DDF[1200](3)}
\,.
\end{align}
For \eqref{eq:12},
the coefficient vanishes once the explicit expression is plugged in.
So this relation does not give any constraint on the amplitudes.
The other relations impose nontrivial relations among DDF amplitudes of a fixed
order of $s$.
So far, there appear nine amplitudes and there are five relations.
The first four relations are simplified as
\begin{align}
{\cal T}^{T_R|T_R}_{DDF[1200](4)} 
=& -\frac{2\hat{t}+1}{\sqrt{(-\hat{t})(1+\hat{t})}} {\cal T}^{T_R|T_RT_R}_{DDF[1200](9/2)}
  \,,\\
{\cal T}^{I_R|I_R}_{DDF[1200](3)} =& \frac{-2(2\hat{t}+1)}{\sqrt{(-\hat{t})(1+\hat{t})}}{\cal T}^{I_R|T_RI_R}_{DDF[1200](7/2)} 
\,,\\
{\cal T}^{T_R|I_RI_R}_{DDF[1200](7/2)}  =& -\frac{1}{2} {\cal T}^{T_R|T_RT_R}_{DDF[1200](9/2)}
\,,\qquad
{\cal T}^{I_R|T_RI_R}_{DDF[1200](7/2)} =
-\mathcal{T}^{TTT}_{DDF[3000](9/2)}
\,.
\end{align}
The relations in the last line are the ones mentioned in Section \ref{sec:high-energy-linear}.
Together with the last relation involving $J_i$ index, one can see that the subleading amplitudes
are related to the higher order part of the amplitudes.
By use of the rotational symmetry, $I_R \rightarrow J$,
one further finds
\begin{align}
  {\cal T}^{I_R|I_R}_{DDF[1200](3)} = 2\mathcal{T}^{T;T}_{DDF[3000](4)} \,. 
\end{align}
Thus, these amplitudes are represented by two of the leading order parts,
say ${\cal T}^{T_R|T_RT_R}_{DDF[1200](9/2)}$ and $\mathcal{T}^{TTT}_{DDF[3000](9/2)}$.
The next order relations involve new leading order part (like $\mathcal{T}^{T}_{DDF[3000](7/2)}$)
or subleading parts of the amplitudes.
In this way, the exact bracket relations provide constraints on
the high-energy expansions
of the DDF amplitudes, and relate them in a complicated manner.
The relation between the leading order parts of the amplitudes
of $T$-projection,
\begin{align}
\mathcal{T}^{TTT}_{DDF[3000](9/2)}=  {\cal T}^{T_R|T_RT_R}_{DDF[1200](9/2)} 
\label{eq:3000-1200_DDF_stringbit1}
\,,
\end{align}
is not obtained also in this case.
If we assume this relation, one can see that the higher order amplitudes
can be written in terms of one of the leading order parts.

%9
\section{Summary and conclusion}
\label{sec:conclusion}

In this paper, through a detailed study of bracket state spectrum and a new definition of $q$-orthonormal helicity basis, we re-examine the exact symmetry 
identities in bosonic open string theory as first derived by G. W. Moore. Based on two illustrative case studies, we are able to spell out the concrete 
kinematic contents of the symmetry relations among tree-level four-point amplitudes of low-lying stringy excitations. These relations are also recast into 
identities among conventional and DDF amplitudes, where the participant states in the scattering amplitudes in each basis are more familiar and form 
well-defined representations of other symmetry groups. In so doing, we can also connect and compare with other well-known symmetry patterns of 
scattering amplitudes in string theory, in particular, the high-energy symmetry as advocated by D. J. Gross. We show that, under certain presumptions, 
part of high-energy symmetry (especially the linear relations as derived from decoupling of the high-energy zero-norm states) can be extracted 
from a high-energy expansion of Moore's symmetry relations. Furthermore, we can detect some of the energy hierarchy of the DDF amplitudes and give 
new inter-level and subleading relations from a high-energy expansion point of view.

To summarize, our findings in this paper are: 

(1) While Moore's exact symmetry identities lead to infinitely
many strong constrains among inter-level amplitudes, their high-energy
limits do not have complete overlap with the high-energy linear
relations as derived from decoupling of high-energy zero-norm states
\cite{Chan:2003ee, Chan:2004yz}.
To make connections and comparisons with the high-energy symmetry of
Gross, we need to make high-energy expansions of both
transformation matrices and the relevant scattering amplitudes. Only
if we obtain a closed set of linear equations among leading components
of scattering amplitudes with various physical polarizations, we can
derive some constraints among leading amplitudes at the same mass
level. Given that the transformation matrices among stringy state
bases are energy-dependent, the mixing of different components
(organized by powers of $s$) of various stringy scattering amplitudes
is unavoidable and the existence of closed relations is definitely
non-trivial.
For instance, \eqref{eq:3000-1200_cov_s4_1} as derived from the order
 $s^{4}$ relations in $\mathcal{A}[\tilde{3}000] =
 \mathcal{A}[1\tilde{2}00]$,
is a constraint among amplitudes,
$\mathcal{T}_{[3000](4)}^{TTL}$, $\mathcal{T}_{[3000](4)}^{LLL}$,
and $\mathcal{T}_{[1200](3)}^{J|JL_2}$,
but it is not strong enough to give
the linear relation as shown in \cite{Chan:2003ee, Chan:2004yz}.
It is, however, not a problem of Moore's relation.
As we have seen, there are infinitely many different ways to realize
bracket operators at a given level, and they are related to
different sets of amplitudes through Moore's relation.
Therefore, the missing link observed here should be connected if we
include further sets of relations.
The failure here means that our simple example is
not a sufficient set to obtain all the known relations.
Note that the successful case of $\mathcal{A}[\tilde{3}000] =
 \mathcal{A}[1\tilde{2}00]$, \eqref{Cov3000-1200_Rel2},
where the linear relation among $\mathcal{T}^{TTT}_{[3000](9/2)}, \;
\mathcal{T}^{TLL}_{ [3000](9/2)}$, and $\mathcal{T}^{[L,T]} _{
  [3000](9/2)}$ can be derived from the high-energy expansions of
Moore's exact identities, in our opinion, should be taken as an
accident.

(2) In our previous studies of the symmetry patterns of stringy
scattering amplitudes \cite{DDF-HE}, we have observed a special feature of the
all-transversely-polarized (with respect to each own momentum)
scattering amplitudes. The high-energy limits of this class of stringy
scattering amplitudes demonstrate some partonic behaviors which we
call string bit pictures. For instance,
\eqref{eq:3000-1200_DDF_stringbit1} provides
one simple example, where as long as the total level numbers and
total spins of scattering states are the same 
($3 + 0 + 0 + 0 = 1 + 2 + 0 + 0$ in this case),
 the scattering
amplitudes are always equal (up to sign). 
For
$\mathcal{A}[\tilde{3}000]=\mathcal{A}[1\tilde{2}00]$ example, such an
inter-level symmetry pattern was only observed by explicit
evaluations of stringy scattering amplitudes. 
While this inter-level symmetry pattern is derived
by Moore's exact symmetry in the case of
$\mathcal{A}[\tilde{2}000]=\mathcal{A}[1\tilde{1}00]$,
due to the small number of the amplitudes,
it is unlikely deducible for higher level amplitudes.
The use of such a sting bit picture  helps in
simplifying high-energy relation in  Moore's exact symmetry
identities. Together with the rotational invariance (trading $I$ (or
$I_R$) polarizations into $J$ direction), we are able to represent all
leading-components (up to order $s^{9/2}$) of stringy scattering
amplitudes in the $\mathcal{A}[\tilde{3}000] = 
\mathcal{A}[1\tilde{2}00]$ relation by a single amplitude
$\mathcal{T}^{TTT}_{DDF[3000](9/2)}$.

(3) In our study, it is clear that due to the energy dependence of the transformation matrices, the sub-leading components of different energy orders in various stringy scattering amplitudes will get mixed in the high-energy relations from Moore's identities. In order to make a systematic expansion and compare various components we need to choose a set of reference kinematic variables, which are $s$ and $\hat{t} \equiv t/s$ in our two case studies.
The fact that we have chosen special scattering processes such that, in each case, both sides of the  Moore's exact identity share the same Mandelstam variables, is simply for the sake of convenience. In general, it is not clear that if there are many sets of Mandelstam variables $\{ (s_a,t_a)^4_{a=1} \}$, how to choose the best reference kinematic variables. In addition, since each amplitude defines its own scattering plane and generally we have to compare scattering processes at different scattering angles. One has to be careful in making conclusions about the sub-leading patterns of the high-energy relations in Moore's exact identities. We give a brief illustration of subtleties about the choice of reference kinematic variables in Appendix \ref{sec:fixed-angle}.

Aside from these symmetry relations and the connections among different approaches which are realized by scattering amplitudes as functions of kinematic 
variables, there are other issues worth further exploration: 
\begin{itemize}
\item The algebra of bracket commutator and the origin of symmetry breakdown in string theory. This is a question of fundamental importance in string 
theory. One can easily imagine the bracket algebra relating different stringy excitation should be some kind of residue symmetry after spontaneous breakdown 
of certain (presumably infinite-dimensional) symmetry. See
discussion in \cite{Moore:1993zc}. One of the original
motivations studying the high-energy symmetry is to 
make an analogy of equivalence theorem in electroweak theory. It is not clear if we can view these exact relations (which apparently look kinematic 
dependent) as a non-linear realization of the broken symmetry. 
\item Similarity between constructions of DDF states and that of
bracket states, especially in the $q$-orthonormal base, is of interest. 
As demonstrated in \cite{West:1994my}, the algebraic structure among bracket states may be understood as a kind of Kac-Moody algebra, at least partially.
\item Following the previous line of thought, one might wish to prove other kinematic limits these 
symmetry relations e.g. Regge limits (\cite{Ko:2008ft}) to see if we can obtain other useful patterns or connect with different approaches. 
\item We can make further 
generalizations by choosing different space-time backgrounds, adding supersymmetry, or studying string theory at finite temperature. The 
study of loop amplitudes may require special efforts. 
\item Similar structure and patterns can be captured in the exactly solvable string theory \cite{Witten:1991zd}. 
In either minimal string models or matrix models one may be able to give more mathematical insight to this ultimate question.
\end{itemize}

\section*{Acknowledgement}

The authors thank S.~Hirano,
Pei-Ming Ho, and Xue-Yan Lin
 for valuable discussion and comments.
They also thank
M.~Asano, H.~Kawai, Y.~Kimura, F.~Sugino,
T.~Yoneya,  and  K.~Yoshida
for comments on the presentation based on this work.
C.-T. Chan is supported in part
by National Science Council (NSC) of Taiwan under the
contract No. 99-2112-M-029-001-MY3.
The work of S.~Kawamoto is supported by NSC99--2112--M--029--003--MY3 and NSC102--2811--M--029--001.
The work of D.~Tomino is supported by  NSC100--2811--M--029--002 and  NSC101--2811--M--029--002. 
The authors are also supported in part
by Taiwan String Theory Focus Group in NCTS under
NSC No. 100-2119-M-002-001.

%%%%%%%%%%%%%%%%%%% Appendix %%%%%%%%%%%%%%%%%%%
\appendix

\section{More on bracket states 
and the physical state conditions}
\label{sec:bracket-operators}

In this subsection, we present an interesting
example in which
a deformer operator that does not satisfy physical state conditions
generates physical bracket operators,
by using
bracket operators in
$\mathcal{A}[\tilde{3}000]=\mathcal{A}[1\tilde{2}00]$ relation.
In this appendix, we take
$\alpha'=1/2$ for simplicity.
The physical state conditions for the level 2 bracket operator
$V_{(2)}^\text{br}(\tilde{k}_2,z)$ are
reduced to the conditions on the deformer polarization tensors as
\begin{align}
  \label{eq:1200_br_phys1}
0=&  [\zeta_q \cdot q + \tilde\zeta_q ]_\mu 
+ q_\mu k_2 \cdot  [\zeta_q \cdot q + \tilde\zeta_q ]
\,,\\
\label{eq:1200_br_phys2}
0=& 
\zeta_{q\mu\nu} \eta^{\mu\nu} + 2\tilde\zeta_q \cdot q
+ 6k_2 \cdot  [\zeta_q \cdot q + \tilde\zeta_q ] \,,
\end{align}
with $q^2 =-2$ and $k_2 \cdot q=-1$.
From $q \cdot k_2=-1$,
we have a solution of  \eqref{eq:1200_br_phys1},
\begin{align}
\zeta_{q\mu\nu} q^\nu + \tilde\zeta_{q\mu} = c q_\mu \,,
\qquad c \in \bC
\label{eq:1200_br_phys1_1}
\,.
\end{align}
It is easy to see that $\zeta_{q\mu\nu} q^\nu + \tilde\zeta_{q\mu}$
does not have a component transverse to $q$, 
and this is a general
solution with one parameter $c$.
Plugging this into \eqref{eq:1200_br_phys2},
we find
\begin{align}
    \zeta_{q\mu\nu} \left( \eta^{\mu\nu} -2 q^\mu q^\nu \right) = 10c \,.
\label{eq:1200_br_phys2_1}
\end{align}
When $c=0$, $\zeta_{q\mu\nu}$ and $\tilde\zeta_{q\mu}$ satisfy the
physical state conditions for ${J}_{(2)}(q,w)$.
Thus, $c$ measures a failure of those conditions.
We are interested in whether the conditions
\eqref{eq:1200_br_phys1_1} and \eqref{eq:1200_br_phys2_1}
allow a solution with $c \neq 0$.
To solve these conditions, we introduce a helicity basis with
respect to $q$;
  $e^{P_q}= q/\sqrt{2}$
and transverse orthonormal vectors $e^{T_i}$
($i=1,\cdots,25$).
When $c \neq 0$, solutions are
\begin{align}
c=\frac{1}{10}: \quad  \zeta_{q\mu\nu} =& e^{T_i}_\mu e^{T_i}_\nu \,,
\quad \tilde\zeta_{q\mu} = \frac{\sqrt{2}}{10} e^{P_q}_\mu \,,
\qquad (i=1,\cdots, 25, \,  \, \text{no sum for $i$})
\label{lv2_exotic-sol1}
\\
c=-\frac{1}{2}: \quad  \zeta_{q\mu\nu} =& e^{P_q}_\mu e^{P_q}_\nu \,,
\quad \tilde\zeta_{q\mu} = \frac{1}{\sqrt{2}} e^{P_q}_\mu \,.
\end{align}
The second choice leads to $V_{(2)}^\text{br}=0$ identically,
while for $c=1/10$ one can check that it indeed gives nonvanishing
bracket operators.
Together with the solutions with $c=0$,
these complete the conditions for the level 2 bracket operator to be physical.

We consider the physical state condition of the level 3
bracket
operator $V_{(3)}^\text{br}(\tilde{k}_1,z)$.
In this case, three polarization tensors provide three physical state
conditions for $\zeta_{1\mu}, \zeta_{q\mu\nu}, \tilde\zeta_{q\mu}$.
These conditions turn out to be too complicated to find a general solution.
However, 
curiously, if we take $V_{(2)}^\text{br}$ case solution \eqref{eq:1200_br_phys1_1}
and the physical state condition for the seed operator $\zeta_1 \cdot
k_1=0$ as an ansatz,
the physical state conditions
lead to a solution which is exactly the same as
\eqref{lv2_exotic-sol1} (with the same $c=1/10$).
It should not be accidental since these two operators are related
through Moore's relation and they would define physical amplitudes
for the same choice of $J_{(2)}$.
It is not a complete analysis, and
there might be other physical choices for $V_{(3)}^\text{br}$,
but we do not pursue it further and stop here.

Thus,
a lesson from these examples is
that it is indeed possible to define physical 
 bracket operators
by using  unphysical  deformer operators
and physical seed operators.
Since an extra physical state in this example
seems very special, 
in the main part, we concentrate on ``standard'' physical choices
where both seed and deformer operators are physical.

\section{DDF states}
\label{sec:ddf-states}

In this appendix, we summarize the basic facts on 
Del Giudice, Di Vecchia, and Fubini (DDF) operators \cite{Del Giudice:1971fp}
and corresponding states.
Our treatment  follows closely that of \cite{Green:1987sp},
 and we will spell out explicit formulas
which are used in the analysis in the main part.

\paragraph{Construction}
\label{sec:construction}

Let $\ket{0;p_0}$ be a tachyonic ground state of the bosonic open string theory
with $p_0^2=1/\alpha'$.
We introduce a null vector $k_0$ which satisfies $k_0^2=0$ and $p_0
\cdot k_0 = 1/(2\alpha')$.
It is straightforward to see that $p_{(N)}=p_0-Nk_0$ satisfies the
mass-shell condition of level $N$ states,
$p_{(N)}^2=(1-N)/\alpha'$.
A convenient parametrization of these momenta are
\begin{align}
\label{eq:DDF_param}
    p_0^\mu =& \frac{1}{\sqrt{\alpha'}} 
\left( 0, 0, \cdots , 1 \right) \,,
\quad
k_0^\mu = \frac{1}{2\sqrt{\alpha'}}
\left( -1, 0, \cdots , 1 \right) \,.
\end{align}
The DDF operator is defined as
\begin{align}
  A_n^\ell(nk_0) =& \oint \frac{dz}{2\pi i} \, \frac{i \partial X^\ell(z)}{\sqrt{2\alpha'}}
e^{i nk_0 \cdot X(z)} \,,
\end{align}
where $z=e^{i\tau }$,
and
$\ell$ refers to the transverse directions with respect
to $p_0$ and $k_0$, $\ell=1,\cdots, 24$.
DDF states are defined by the action of $A_{-n}^\ell$ on
the tachyonic ground state $\ket{0;p_0}$.
Since $p_0 \cdot k_0=1/(2\alpha')$, the action of a
DDF operator on the tachyonic ground state
$\ket{0;p_0}$ is well-defined.
$A_n^i$ commutes with Virasoro operators $L_n$ and their commutation
relation $[A_n^\ell, A_m^k]= n \delta^{\ell k} \delta_{n+m}$ is the
same as the standard transverse oscillators $\alpha_{-n}^\ell$.
Thus the DDF states generates the whole positive norm physical states.

We are ready to write down DDF states in terms of standard
oscillators $\alpha_n^\mu$.
It is convenient to use the helicity basis where
the inner product with $k_0$ is proportional to $L-P$ projection,
$k_0 \cdot \alpha_{-n} \propto (e^L-e^P)_\mu \alpha_{-n}^\mu =
\alpha_{-n}^{(L-P)}$ \cite{Chan:2005ji}.
For the first few levels, the result is
\begin{align}
\text{level 1}:  \quad &
\ket{a;p_{(1)}}_{DDF} \equiv    A_{-1}^a \ket{0; p_0} 
=
\alpha_{-1}^a
 \ket{0; p_{(1)}}  \,,
\\
\text{level 2}:  \quad &
\ket{ab;p_{(2)}}_{DDF} \equiv 
A^a_{-1} A^b_{-1} \ket{0; p_0}
\nn\\&\hskip2em
=
\left[
\alpha_{-1}^{ab}
+\delta^{ab}
\left(
-\frac{1}{2\sqrt{2}} \alpha_{-2}^{(L-P)}
+\frac{1}{4} \alpha_{-1}^{(L-P)} \alpha_{-1}^{(L-P)}
\right)
\right] \ket{0;p_{(2)}}
\,,
\label{eq:DDF2ij}
\\
&
\ket{a;p_{(2)}}_{DDF} \equiv   A_{-2}^a \ket{0;p_0}
=
\left(
\alpha_{-2}^a
- \sqrt{2} \alpha_{-1}^a \alpha_{-1}^{(L-P)}
\right) \ket{0;p_{(2)}}
\label{eq:DDF2i}
\,,
\end{align}
\begin{align}
&
\text{level 3}:
\nn\\
&
\ket{abc;p_{(3)}}_{DDF} \equiv 
A^a_{-1} A^b_{-1}  A_{-1}^c \ket{0; p_0}
\nn\\ &
=
\bigg[
\alpha_{-1}^{abc}
+ \bigg( \delta^{ab}\alpha_{-1}^c  + \delta^{bc}\alpha_{-1}^a+
 \delta^{ca}\alpha_{-1}^i
\bigg)
\left(
-\frac{1}{4} \alpha_{-2}^{(L-P)}
+\frac{1}{8} \alpha_{-1}^{(L-P)} \alpha_{-1}^{(L-P)}
\right)
\bigg] \ket{0;p_{(3)}}
\,,\\
&
\ket{a;b \, ;p_{(3)}}_{DDF} \equiv
A^a_{-2} A^b_{-1} \ket{0; p_0}
\nn\\&=
\left[
\alpha_{-2}^{a} \alpha_{-1}^b
-\alpha_{-1}^{ab} \alpha_{-1}^{(L-P)}
\right.\nn\\&\left.\hskip3em
+\delta^{ab}
\left(
-\frac{1}{3} \alpha_{-3}^{(L-P)}
+\frac{1}{2} \alpha_{-2}^{(L-P)} \alpha_{-1}^{(L-P)}
-\frac{1}{6} \left( \alpha_{-1}^{(L-P)} \right)^3
\right)
\right] \ket{0;p_{(3)}}
\,,\\
&
\ket{a;p_{(3)}}_{DDF} \equiv
   A_{-3}^a \ket{0;p_0}
\nn\\=&
\left[
\alpha_{-3}^a
- \frac{3}{2} \alpha_{-2}^a \alpha_{-1}^{(L-P)}
+ \alpha_{-1}^a
  \left(
  -\frac{3}{4} \alpha_{-2}^{(L-P)}
  +\frac{9}{8} \left( \alpha_{-1}^{(L-P)} \right)^2
 \right)
\right] \ket{0;p_{(3)}}
\,.
\end{align}
When it is obvious, the momenta may not be displayed and the 
states are expressed by its transverse indices.
It should be noted that the definition of $e^P$ and $e^L$ depends on $p_{(N)}$
and they are different for each level.
At level 1, there is no
distinction
between DDF states and usual massless states.

\paragraph{From positive norm states to DDF states}
\label{sec:from-covar-posit}

DDF states form a basis of physical states at a given level
and then
it is possible to rewrite a given physical state
by use of them  up to zero-norm states\footnote{The structure of
  zero-norm states in terms of the helicity basis is discussed in
  \cite{Chan:2005qd}.}.
We utilize such decomposition to relate level 2 and 3 
bracket states to DDF states.

\subparagraph{Level 2}
\label{sec:level-2}
First, we list the zero-norm states at level 2,
\begin{align}
\ket{\text{ZN}_1} =&    \left(
5\alpha_{-1}^{PP}
+\alpha_{-1}^{LL}
+\sum_{a=1,\cdots,24} \alpha_{-1}^{aa}
+5\sqrt{2} \alpha_{-2}^P
\right) \ket{0;k}
\label{eq:ZN_lv2_2}
\,, \\
\ket{\text{ZN}_2} =&    \left(
\sqrt{2} \alpha_{-1}^{PL}
+ \alpha_{-2}^L
\right) \ket{0;k}
\label{eq:ZN_lv2_1L}
\,, \\
\ket{\text{ZN}_{3}^a} =&    \left(
\sqrt{2} \alpha_{-1}^{Pa}
+ \alpha_{-2}^{a}
\right) \ket{0;k}
\label{eq:ZN_lv2_1a}
\,,
\end{align}
where $e^P$ and $e^L$ represent the momentum and the 
longitudinal helicity with respect to the momentum $k$.
$e^a$ ($a=1,\cdots,24$) represents the transverse directions
with respect to $k$.
We now consider a decomposition of the following level 2 physical
positive norm state,
\begin{align}
  \label{eq:gen_lev2_cov_posi}
&  \left[
\sum_{a',b'} G_{a'b'} \alpha_{-1}^{a'b'}
+G \sum_{a'} \alpha_{-1}^{a'a'}
\right] \ket{0;k}
\,,
\end{align}
where $a',b'$ indices run over $L, a$,
the transverse directions together with the longitudinal
directions,
and the physical state conditions imply
  $\sum_a G_{aa} + G_{LL} + 25 G =0$.
This state can indeed be written in terms of the DDF states,
up to zero norm states, as
\begin{align}
  \eqref{eq:gen_lev2_cov_posi}
=&
\sum_{a,b} D_{ab} \ket{ab;k}_{DDF}
+\sum_a D_a \ket{a;k}_{DDF}
+D \sum_a \ket{aa;k}_{DDF}
% \nn\\&
% +(\text{zero-norm states}) 
\,,
\label{eq:gen_lev2_DDF}
\\
&
D_{ab}= G_{ab}
\,,
\qquad
D_a=
-\sqrt{2}G_{L a}
\,,\qquad
D=
\frac{1}{4} \big( G_{LL}  + 5 G \big)
\,.
\end{align}

\subparagraph{Level 3}
\label{sec:level-3}

The zero-norm states of this level are
\begin{align}
    \ket{ZN_1^{aa}} =&
\left(
\alpha_{-1}^{aa P}
- \alpha_{-1}^{LLP}
+ \alpha_{-2}^{a} \alpha_{-1}^{a}
- \alpha_{-2}^L \alpha_{-1}^{L}
\right) \ket{0;k}
\,,\\
    \ket{ZN_2^{a'b'}} =&
\left(
\alpha_{-1}^{a'b' P}
+\alpha_{-2}^{(a'} \alpha_{-1}^{b')}
\right) \ket{0;k}
\,,
\qquad (a' \neq b')
\\
    \ket{ZN_3^{a'}} =&
\left(
9 \alpha_{-1}^{PPa'}
+\alpha_{-1}^{LLa'}
+\sum_{b} \alpha_{-1}^{a'bb}
+18 \alpha_{-2}^{(P} \alpha_{-1}^{a')}
+6 \alpha_{-3}^{a'}
\right) \ket{0;k}
\,,\\
    \ket{ZN_4^{a'}} =&
\left(
\alpha_{-1}^{LLa'}
+\sum_{b} \alpha_{-1}^{a'bb}
+9 \alpha_{-2}^{[P} \alpha_{-1}^{a']}
-3 \alpha_{-3}^{a'}
\right) \ket{0;k}
\,,\\
\ket{ZN_5} =&
\left(
25 \alpha_{-1}^{PPP}
+75 \alpha_{-2}^P \alpha_{-1}^P
+50 \alpha_{-3}^P
\ph{\frac{1}{2}}
%\right.\nn\\&\left.\hskip1em
+9\sum_{b} \left(
 \alpha_{-1}^{bb P} + \alpha_{-2}^{b} \alpha_{-1}^{b}
 \right)
\right) \ket{0; k} \,,
\end{align}
where $a,b,c=1,\cdots,24$
are transverse directions with respect to $k$
and $a'=a, L$.
The indices are not summed over otherwise explicitly displayed.
As in the level 2 case, we
consider a decomposition of the following physical 
positive norm state in terms of DDF states,
\begin{align}
  \label{eq:gen_lev3_cov}
&
  \bigg[
\sum_{a',b',c'} G_{a'b'c'} \alpha_{-1}^{a'b'c'}
+\sum_{a',b'} G_{[a'b']} \alpha_{-2}^{[a'} \alpha_{-1}^{b']}
\bigg] \ket{0; {k}} 
\,,
\end{align}
with the physical state conditions,
$\sum_b G_{abb} + G_{aLL}=0$ and 
$\sum_b G_{Lbb} +  G_{LLL} =0$.
After some algebra, we find that 
\eqref{eq:gen_lev3_cov} can be rewritten as
\begin{align}
&
\sum_{a,b,c} D_{abc} \ket{abc; k}_{DDF}
+\sum_{a,b} D_{(ab)} \ket{(a;b) \, ; k}_{DDF}
+\sum_{a,b} D_{[ab]} \ket{[a;b] \, ; k}_{DDF}
\nn\\&
+\sum_a D_{1\, a} \ket{a \, ; k}_{DDF}
+\sum_a D_{2\, a} \sum_b \ket{abb \, ; k}_{DDF}
+D \sum_b \ket{b;b \, ; k}_{DDF}
% \nn\\&
% +(\text{zero-norm states})
\,,
\end{align}
up to zero norm states, and
\begin{align}
  D_{abc}=& G_{abc} 
\,,\qquad
D_{(ab)} =-3G_{Lab} \,,
\qquad
D_{[ab]} = G_{[ab]} \,,
\qquad
D = -\frac{1}{2}G_{LLL} 
\nn\\
D_{1a} =& \frac{1}{4} \left( 9 G_{LLa} +2 G_{La} \right)
\,,\qquad
 D_{2a} = \frac{1}{8} \left( 3 G_{LLa} - 2 G_{La} \right)
\,.
\end{align}

\section{High-energy expansion with a fixed scattering angle}
\label{sec:fixed-angle}

In this paper, we focus on the high-energy limits
($s\rightarrow \infty$ 
with $\hat{t}=t/s$ fixed) of string scattering amplitudes.
In the leading order,
this limit corresponds to the fixed-angle high-energy limit
of the amplitudes,
but when we consider relation
among subleading part of the amplitudes, there appears some difference.
In this appendix, we discuss the high-energy asymptotic relation
with the fixed scattering angle by taking
$\mathcal{A}[\tilde{2}000]=\mathcal{A}[1\tilde{1}00]$ as an example.

Recall that Mandelstam variables $s$ and $t$ are common
on both hands sides in this example.
Due to the mass difference, $t$ takes different forms, in terms of
$s$ and the scattering angles, on both hand sides as
\begin{align}
    t_{[\tilde{2}000]}=&
 -\frac{s+4}{2} + \frac{\sqrt{(s^2+16)(s+8)}}{2\sqrt{s}}\cos\theta \,,
\label{eq:t_2000}
\\
t_{[1\tilde{1}00]}=&  -\frac{s+4}{2} + \frac{\sqrt{s(s+8)}}{2}\cos\theta' \,,
\label{eq:t_1100}
\end{align}
where $\theta$ is the scattering angle for $\mathcal{A}[\tilde{2}000]$
and $\theta'$ for $\mathcal{A}[1\tilde{1}00]$,
and we have set $\alpha'=1/2$.
Since actually $t_{[\tilde{2}000]}=t_{[1\tilde{1}00]}$, these two
angles are related as
$\cos\theta' = \cos\theta \big( 1+ 8s^{-2}-32 s^{-4}+\cdots \big)$.
$\hat{t}$ and $\theta$ are related as 
$\hat{t}=-\sin^2 \frac{\theta}{2}+\mathcal{O}(s^{-1})$ in the
high-energy limit.
At the leading order, the expressions are the same
for $\theta$ and $\theta'$, but subleading corrections take
different forms for them.

This poses a question on the high-energy expansions of string
scattering amplitudes.
We need to choose which angle to be fixed under
the $s\rightarrow\infty$ limit.
In this sense, a fixed-angle high-energy limit is ambiguous in
the bracket relation.
To proceed the analysis, we now choose $\theta$ to be fixed
and expand the coefficients and the DDF amplitudes
with $\theta$ fixed,
\begin{align}
    {\cal T}_{DDF[2000]}^{TT}=& 
{\cal T}_{\theta[2000]\; (3)}^{TT}\, s^3 
+\cdots
\,,\quad
  D^{I_q I_q}_{II} =
D^{I_q I_q}_{\theta\, II \; (0)}
 + D^{I_q I_q}_{\theta\, II \; (-1)} \, s^{-1}
+\cdots \,,
\end{align}
where each coefficient is now a function of $\theta$;
for example, ${\cal T}_{\theta[2000]\; (3)}^{TT}=\frac{1}{16}\sin^2\theta$.
It should be emphasized that, apart from the leading ones like
${\cal T}_{\theta[2000]\; (3)}^{TT}$ or $D^{I_q I_q}_{\theta\, II \; (0)}$,
the coefficients are not simply obtained by identifying
$\hat{t}=-\sin^2 \frac{\theta}{2}$ in the fixed-$\hat{t}$ coefficients
that appeared in Section \ref{sec:HE_1200-1100_DDF}.
At $\mathcal{O}(s^3)$
and
at $\mathcal{O}(s^2)$ with $(A,B)=(T_q,T_q), (I_R,I_R), (J,J)$
the relations take the same forms as
\eqref{eq:2000-1100_DDF_3}--\eqref{eq:2000-1100_DDF_II_2}.
For $(A,B)=(T_q,I_q)$ at $\mathcal{O}(s^2)$, we find
\begin{align}
  2\,\mathcal{T}^{TT}_{\theta[2000](3)}=\tan\theta \,
 \mathcal{T}^T_{\theta[2000](5/2)} \,.
\end{align}
This relation involves only in leading amplitudes
of $\mathcal{A}[\tilde{2}000]$ side
and reproduces a known fixed-$\theta$ relation.
At $\mathcal{O}(s)$, after simplifying
by use of higher order relations and rotational symmetry,
$(A,B)=(T_q,T_q), (I_q,I_q), (T_q, I_q)$ and $(J,J)$ choices lead to,
in order,
\begin{align}
-2\,  \mathcal{T}^{T_R|T_R}_{\theta[1100](1)}=&
8\,\cot\theta \mathcal{T}^T_{\theta[2000](5/2)} 
+ \mathcal{T}^{TT}_{\theta[2000](2)}
+19 \mathcal{T}^{II}_{\theta[2000](2)}
%\\&
-2\,
\mathcal{T}^{TT}_{\theta[2000](1)}
\,,\nn\\
2\,\mathcal{T}^{I_R|I_R}_{\theta[1100](1)}=&
8\,\cot\theta\mathcal{T}^T_{\theta[2000](5/2)}\,
-57\,   \mathcal{T}^{II}_{\theta[2000](2)}-2\,   \mathcal{T}^{II}_{\theta[2000](1)}
%\nn\\&
-3\,   \mathcal{T}^{TT}_{\theta[2000](2)}
\,,\nn\\
0=&
 \tan\theta \left(4 \mathcal{T}^T_{\theta[2000](5/2)}
+ \mathcal{T}^T_{\theta[2000](3/2)} \right)\, 
-2  \mathcal{T}^{TT}_{\theta[2000](2)}+2  \mathcal{T}^{II}_{\theta[2000](2)}
\,,\nn\\
   2\,\mathcal{T}^{J|J}_{\theta[1100](1)}\,=&
8\,\cot \theta
\mathcal{T}^T_{\theta[2000](5/2)}
-19\,\mathcal{T}^{II}_{\theta[2000](2)}\,
+2\,\mathcal{T}^{II}_{\theta[2000](1)}\,
-\mathcal{T}^{TT}_{\theta[2000](2)}\,   
\,.\nn
\end{align}
By solving them, one can find several relations among the amplitudes.
However, for subleading amplitudes on $\mathcal{A}[1\tilde{1}00]$
side,
such as $\mathcal{T}^{T_R|T_R}_{\theta[1100](1)}$, this expansion is
not of direct physical relevance
and it is not clear how useful these expressions are.

%%%%%%%%%%%%%%%%%%

\end{document}